\documentclass[twocolumn,twocolappendix,trackchanges]{aastex63}
\usepackage{amsmath}
\newcommand{\code}[1]{\texttt{#1}}

\newcommand{\MESA}{\code{MESA}}

\newcommand{\kms}{{\mathrm{km\ s^{-1}}}}

\usepackage{CJK}

\DeclareRobustCommand{\Figref}[1]{Fig.~\ref{#1}}
\DeclareRobustCommand{\Tabref}[1]{Tab.~\ref{#1}}
\DeclareRobustCommand{\Secref}[1]{Sec.~\ref{#1}}
\newcommand{\zoph}{$\zeta$ Oph}

\newcommand{\Msun}{\ensuremath{\,M_\odot}}

\newcommand{\referee}[1]{#1}
\defcitealias{villamariz:05}{VH05}

\begin{document}

\graphicspath{{./figures/}}

\title{Evolution of accretor stars in massive binaries: broader
  implications from modeling $\zeta$ Ophiuchi}

\author[0000-0002-6718-9472]{M.~Renzo}
\affiliation{Department of Physics, Columbia University, New York, NY 10027, USA}
\affiliation{Center for Computational Astrophysics, Flatiron Institute, New York, NY 10010, USA}

\author[0000-0002-6960-6911]{Y.~G\"otberg}\thanks{Hubble Fellow}
\affiliation{The Observatories of the Carnegie Institution for
  Science, 813 Santa Barbara Street, Pasadena, CA 91101, USA}

\begin{abstract}
  Most massive stars are born in binaries close enough for mass
  transfer episodes. These modify the appearance, structure, and
  future evolution of both stars. We compute the evolution of a
  100-day period binary consisting initially of a 25\Msun\ star and a
  17\Msun\ star, which experiences stable mass transfer. We focus on
  the impact of mass accretion on the surface composition, internal
  rotation, and structure of the accretor. To anchor our models, we
  show that our accretor broadly reproduces the
  properties of $\zeta$ Ophiuchi, which has long been proposed to have
  accreted mass before being ejected as a runaway star when the
  companion exploded. We compare our accretor to models of
  single rotating stars and find that the later and stronger spin-up
  provided by mass accretion produces significant
  differences. Specifically, the core of the accretor retains higher
  spin at the end of the main sequence, and a convective layer
  develops that changes its density profile. Moreover, the surface of
  the accretor star is polluted by CNO-processed material donated by
  the companion. Our models show effects of mass accretion in
  binaries that are not captured in single rotating stellar
  models. This possibly impacts the further evolution (either in a
  binary or as single stars), the final collapse, and the resulting
  spin of the compact object.
\end{abstract}


\section{Introduction}
\label{sec:intro}

The overwhelming majority of massive stars is born in multiple systems
\citep[e.g.,][]{mason:09, almeida:17, moe:17}, and a large fraction will
exchange mass or merge with a companion in their lifetime
\citep[e.g.,][]{sana:12}. The most common type of interaction is
stable mass transfer through Roche lobe overflow (RLOF) after the end of the donor star's main sequence (case
B, \citealt{kippenhahn:67}), when stars experience most of their
radial expansion.  Many
studies have focused on the dramatic impact that these interactions have on the
donor star \citep[e.g.,][]{morton:60, yoon:17, gotberg:17, gotberg:18, laplace:20,
  laplace:21, blagorodnova:21}. Often the accreting companion is treated as a point mass.
However, binary interactions have a crucial impact on the initially
less massive star too.

\subsection{The importance of accretor stars}

During mass transfer, the initially less massive star is expected to
accrete mass, spin-up to critical rotation \citep[e.g.,][]{packet:81},
and possibly be polluted by nuclearly processed material from the
inner layers of the donor star \citep[e.g.,][]{blaauw:93}. The
increase in mass leads to growth of
the convective core and consequent
``rejuvenation'' of the accretor star \citep[e.g.,][]{neo:77,
  schneider:16}. Understanding the evolution of accretor stars in massive
binaries has wide and crucial implications for stellar populations,
electromagnetic transient observations, and gravitational-wave
progenitors.

Accretor stars can appear as blue stragglers
\citep[e.g.,][]{chen:09, chen:10, rain:21} and thus impact cluster
populations, their age estimates, and their main sequence
\citep[e.g.,][]{pols_marinus:94, wang:20} and post main sequence morphology
\citep[e.g.,][]{wei:21}. The high spin of the
accretor star after mass transfer might be the dominant explanation for the origin of
Oe and Be stars (i.e., hydrogen-rich stars showing emission lines,
e.g., \citealt{pols:91}, \citealt{bodensteiner:20},
\citealt{vinciguerra:20}, \citealt{dorigo-jones:20}, \citealt{wang:21_sdOBe}, \citealt{hastings:21}, see
\citealt{rivinius:13} for a review). Rotationally-mixed accretor
stars at low metallicity have also been suggested to be important for the ionizing flux of high-redshift
galaxies \citep[e.g.,][]{eldridge:12, stanway:16}.

After mass transfer, the majority of massive binaries will be
disrupted by the first supernova, and eject the accretor star \citep[``binary SN
scenario'', ][]{blaauw:61, dedonder:97, eldridge:11, boubert:18,
  renzo:19walk, evans:20}.  A small fraction of these would be
sufficiently fast to become runaway stars, but the majority will be
too slow to stand out in astrometric surveys. Assuming a constant star
formation history, \cite{renzo:19walk} estimated that
$10.1^{+4.6}_{-8.6}\%$ ($0.5^{+2.1}_{-0.5}\%$) of O-type stars could
be slow ``walkaway'' (runaway) accretor stars ejected after the companion's
core collapse. Therefore, presently single O-type stars, that accreted
mass earlier on, contribute to the populations of field massive stars \citep[e.g.,][]{dorigo-jones:20}.

Massive accretor stars are also
important from the transients perspective: \cite{zapartas:19} estimated that $14_{-11}^{+4}\%$ of
hydrogen (H) rich
type II SNe come from progenitors that were ejected from a binary. The fact that they accreted mass before exploding can
influence their helium (He) core mass and thus the explosion
properties and the inferred progenitors \citep{zapartas:21}. 
The high post-mass transfer rotation
rate of accretor stars in binaries might have implications for the
formation of long gamma-ray burst progenitors \citep[e.g.,][]{cantiello:07}.

The majority of isolated binary evolution scenarios for
gravitational-wave progenitors include a common-envelope
phase during which a compact object is engulfed in the envelope of a star. This phase is initiated by the initially less massive star, which
accreted mass from its companion before the formation of the first compact object
\citep[e.g.,][]{belczynski:16nat, tauris:17,
  broekgaarden:21}. That accretion of mass,
before the formation of the first compact object, could significantly modify the
internal structure of the star that will initiate the common-envelope
\citep[e.g.,][]{law-smith:20, klencki:21}. Specifically, the
rotation rate, chemical composition, and innermost structure of the
envelope (because of rejuvenation) is expected to differ from that of a single star.

Despite their importance, accretor stars in binaries have so far
received relatively little attention, with the pioneering
studies of \cite{ulrich:76, hellings:83, hellings:84}, and
\cite{braun:95} as notable exceptions. Large grids of accretor star models
are lacking and most existing studies focus on lower mass systems
(e.g., $M_1\lesssim 16\,M_\odot$ in \citealt{vanrensbergen:11}) or
neglect the crucial impact of rotation (e.g., \citealt{sravan:19}, but see also
\citealt{wang:20}). Only few massive accretor star models exist
\citep[e.g.,][]{petrovic:05, cantiello:07}.

Modeling accretor stars requires following the coupled evolution of
two rotating stars that exchange mass. Making robust predictions is
challenging because of the large number of uncertain parameters that are necessary
to model each individual star and their interactions. Population
synthesis calculations based on single star models cannot include the
effects of binary mass transfer on the internal structure, and rely on
the implicit assumption that the accretor is sufficiently well
described by a (possibly fast-rotating) single star model. Here, we
compute detailed evolutionary models of both stars in a binary and
compare our accretor star to rotating single star models to test these
assumptions.

\subsection{A prototypical example of an accretor: $\zeta$ Ophiuchi}

The nearest O-type star to Earth, $\zeta$
Ophiuchi 
(\zoph) provides the opportunity to constrain massive binary evolution
models, as it is believed to be the ejected accretor star from a binary system.
 \zoph\ is located $107\pm4$\,pc from Earth
\citep[][and references therein]{neuhauser:20}, and has the spectral type
O9.5{\rm IVnn} \citep{sota:14}. It occasionally shows emission lines,
making it an Oe star \citep{walker:79, vink:09}. Its surface rotation
rate is extremely high, with most estimates of the projected
rotational velocity from optical spectra exceeding
$v\sin(i)\gtrsim 400\,\kms$ (corresponding to the ``nn'' in the
spectral type, \citealt{zehe:18} and references therein). By comparing
the observed $v\sin(i)=432\pm16\,\kms$ to the theoretical breakup
rotation, \cite{zehe:18} constrained the inclination angle to
$i\gtrsim 56$\,degrees. Using optical interferometry, \cite{gordon:18}
measured the centrifugal distortion of \zoph, finding a polar radius
of $7.5\,R_\odot$ and centrifugally increased equatorial radius of
$9.1\,R_\odot$, corresponding to a $v\sin(i)=348\,\kms$.

\zoph\ was originally
identified as a runaway because of its large proper motion by
\cite{blaauw:52}. Unfortunately, the \emph{Gaia} data for this object
are not of sufficient quality\footnote{The renormalized unit weighted
  error (RUWE) of this star in Gaia EDR3 is 4.48.} to improve previous astrometric results,
but estimates of the peculiar velocity range in $20-50\,\kms$
\citep[e.g.,][]{zehe:18, neuhauser:20}. The large velocity with
respect to the surrounding interstellar material is also confirmed by the
presence of a prominent bow-shock \citep[e.g.,][]{bodensteiner:18}.

Because of its young apparent age, extremely fast rotation, and
nitrogen (N) and He rich surface \citep[e.g.,][]{herrero:92,
  blaauw:93, villamariz:05, marcolino:09}, \zoph\ is a prime example
of runaway from the binary SN scenario \citep{blaauw:93}. Many studies
have suggested that \zoph\ accreted mass from a companion before
acquiring its large velocity, both from spectroscopic and kinematic
considerations \citep[e.g.,][]{blaauw:93, hoogerwerf:00,
  hoogerwerf:01, tetzlaff:10, neuhauser:20} and using stellar modeling
arguments \citep[e.g.,][]{vanrensbergen:96}. Recently,
\cite{neuhauser:20} suggested that a supernova in
Upper-Centaurus-Lupus that occurred $1.78\pm0.21$\,Myr ago produced the pulsar PSR B1706-16, ejected the companion \zoph,
and also injected the short-lived radioactive isotope
$^{60}\mathrm{Fe}$ on Earth. This argues strongly for a successful SN
explosion of the companion with a $\sim 250\,\kms$ natal kick,
sufficient in most cases to disrupt the binary
\citep[e.g.,][]{tauris:15, renzo:19walk, evans:20}.

Although the nature of \zoph\ as a binary product is well established, previous attempts
to model it rely purely on mixing due to rotation as explanation the N- and
He-rich surface composition, because of its observed large surface rotation rate \cite[e.g.,][]{maeder:00}. Also in binary models, chemical enrichment on the surface has been assumed to originate from the spin-up triggering rotational mixing which dredges up processed material from the stellar core (\citealt{vanrensbergen:96}, see also \citealt{cantiello:07}). However, \cite{villamariz:05}
(hereafter, \citetalias{villamariz:05}) were unable to obtain a satisfying fit
for the stellar spectra using the single-star rotating models from
\cite{meynet:00, meynet:03}: by the time rotational mixing enriches
the surface, single massive stars have significantly spun down through
wind mass loss.

Models predict lower efficiency of rotational mixing for
metal-rich and relatively low mass stars. The reason is the increased
importance of mean molecular weight gradients and the longer thermal
timescales compared to more massive stars \citep[e.g.,][]{yoon:06,
  perna:14}. The parent association of \zoph, identified by
\cite{neuhauser:20}, has a metallicity $Z=0.01\simeq Z_\odot$
\citep[based on asteroseismology from][]{murphy:21} and mass estimates
for \zoph\ range from $13-25\,M_\odot$. This mass is at the lower end
of the range where rotational mixing is thought to be able to
efficiently bring CNO-processed material to the surface (chemically
homogeneous evolution, \citealt{maeder:00}).
Given the challenges in explaining the surface composition of \zoph\
as a rotating, single star and the evidence for its past as a member of
a binary system, this star offers a unique opportunity to constrain
the evolution of accretors in massive binaries.

Here, we present
self-consistent binary evolution models of the coupled evolution of
both stars and their orbit. After describing our \texttt{MESA} setup
in \Secref{sec:methods}, we present a model which reproduces the
majority of the salient features of \zoph\ in
\Secref{sec:best_model}. We show the binary mass transfer evolution in
\Secref{sec:MT}, before focusing on the accretor's rotational
evolution in \Secref{sec:rot} (compared to single stars in
\Secref{sec:rot_comparison_single}), and its internal mixing processes
in \Secref{sec:mixing} (again compared to single star models in
\Secref{sec:mix_comparison_single}). We discuss the sensitivity of our
results to the many uncrtain parameters in \Secref{sec:discussion}, before
concluding in \Secref{sec:conclusions}.

\section{Modeling massive binaries with \texttt{MESA}}
\label{sec:methods}

We follow the coupled evolution of two massive stars in a binary
system using \texttt{MESA} (version 15140, \citealt{paxton:11,
  paxton:13, paxton:15, paxton:18, paxton:19}). Our choice of input
parameters is available at \url{https://github.com/mathren90/zeta_oph}
and all our numerical results are available at
\url{https://doi.org/10.5281/zenodo.4701565}. We discuss here only the
most relevant parameters and describe the effects of varying some of
the adopted values in
\Secref{sec:discussion}. Appendix~\ref{sec:software} gives more
details on our choice of input physics, and in
appendix~\ref{sec:res_tests} we discuss the numerical resolution.

We adopt the \cite{ledoux:47} criterion to determine convective
stability and a mixing length parameter of $1.5$. We allow for
time-dependent convection as in \cite{renzo:20:ppi_conv} based on
\cite{arnett:69}. We include semiconvection and thermohaline mixing
following \cite{langer:83} and \cite{kippenhahn:80}, respectively,
each with efficiency $1.0$.  We use the exponential core overshooting
from \cite{herwig:00} with free parameters
$(f, f_0)=(4.25\times10^{-2}, 10^{-3})$ \citep{claret:17} which
broadly reproduces the width of the main sequence from \cite{brott:11}
for a 16\Msun\ single star. We do not account for over/undershooting for
off-center convective layers. We also use the local implicit
enhancement of the convective flux in superadiabatic regions
introduced in \texttt{MESA} 15140 (\texttt{use\_superad\_reduction = .true.}).

We treat rotation in the ``shellular'' approximation
\citep[e.g.,][]{zahn:92, ekstrom:12}, that is, we assume constant
rotational frequency $\omega$ along isobaric surfaces. Furthermore, we
assume tidal synchronization and rigid rotation at the beginning of
our runs, at the zero age main sequence (ZAMS). 
Our models include a diffusive
approximation for meridional currents (Eddington-Sweet circulations,
\citealt{sweet:50}), which dominate the chemical mixing due to
rotation. We also include the secular and dynamical shear
instabilities, and the Goldreich-Schubert-Fricke (GSF) instability
(see \cite{heger:00} for a review of these processes).
We assume a
Spruit-Tayler dynamo for the transport of angular momentum
\citep{spruit:02}, and choose the same parameters as
\cite{heger:00}. This also includes the rotational enhancement of the
wind mass loss as in \cite{langer:98}. Moreover, when a star
approaches critical rotation, we re-calculate the wind mass
loss rate implicitly to keep its rotation sub-critical. Specifically, at each
time step we calculate a wind enhancement factor to reach the ratio
$\omega/\omega_\mathrm{crit}\lesssim 0.95$ where $\omega$ is the
angular spin frequency and
$\omega_\mathrm{crit}=\sqrt{(1-L/L_\mathrm{Edd})GM/R^3}$ is the critical spin frequency, with
$L_\mathrm{Edd}$ the Eddington luminosity computed using the stellar
opacity down to optical depth $\tau=2/3$, $L$ is the luminosity, $R$
the radius of the equator, and $G$ the gravitational
constant. 

We assume a metallicity of $Z=0.01$ informed by the present-day $Z$ of
\zoph's parent cluster \citep{murphy:21}. The relative
element abundances are assumed to scale with solar values \citep{grevesse:98}.

We include stellar winds following \cite{vink:00,vink:01} for
effective temperature $T_\mathrm{eff}\geq 11\,000$\,K, unless the
surface H mass fraction is lower than 0.4, when we instead use
the mass-loss rate from \cite{nugis:00}. For $T_\mathrm{eff}\leq 10\,000$\,K, we use the
mass-loss rate from \cite{dejager:88}, and interpolate linearly in
between the hot and cold wind for intermediate $T_\mathrm{eff}$.
We assume a wind scaling factor of 1 \citep{smith:14, renzo:17}.

Compared to the measurements for
\zoph, we might be over-estimating the wind mass-loss rate by a factor
of 100 \citep[``weak wind problem'', see][]{marcolino:09}. This
suggests that spin-down due to wind mass-loss may also be
overestimated compared to reality. However, \cite{lucy:12} and
\cite{lagae:21} proposed that the temperature structure of the winds
of low-luminosity O-type stars might affect the spectral lines and
cause an empirical underestimate of the mass-loss rate.

We follow \cite{kolb:90} to calculate the mass transfer rate with an
implicit scheme. Moreover, we assume that the specific angular
momentum and entropy of the transferred mass match the surface of the
accretor, while the chemical composition is set by the stratification
of the donor star. Mass transfer is conservative until the accretor
reaches critical rotation, after which rotationally-enhanced mass loss
governs the mass transfer efficiency \citep[e.g.,][]{petrovic:05}.
Transferred matter which is not successfully accreted carries away the
specific angular momentum corresponding to the accretor's orbital
motion \citep[e.g.,][]{soberman:97, vandenheuvel:17}.

For simplicity, we evolve the accretor star as single after the mass
transfer is completed. We define this based on the surface properties
of the donor star, specifically, we require that is has lost most of
its hydrogen-rich envelope -- surface He mass fraction exceeds 0.35 --
and its radius is smaller than both its Roche radius and the ZAMS
radius. The stripped donor star is contracting at this point, and
will likely develop strong wind mass loss, possibly appearing as a
Wolf-Rayet star. Depending on the amount of leftover H-rich material
and the wind mass-loss rate, it is possible that the donor would
expand again later in the evolution, filling its Roche lobe anew
\citep[cf.,][]{gilkis:19}. However, this is not expected for the mass
and metallicity we focus on here \citep[e.g.,][]{laplace:20}. We stop the evolution
of the accretor star at terminal age main sequence (TAMS) defined as
when the central mass fraction of H drops below $10^{-4}$.

To compare with our accretor model, we also compute the main sequence
evolution of non-rotating single stars of 15, 17, 20, 25, and
30$\,M_\odot$ and four more single $20\,M_\odot$ stars rigidly rotating with
$\omega/\omega _{\rm crit} = $ 0.2, 0.3, 0.4, and 0.5 at birth. Apart
from the initial rotation rate and the fact that they are single, the
setup of these models is otherwise identical to our stars in the binary system.

\section{Evolution of an accretor star}
\label{sec:best_model}

\begin{figure}[htbp]
  \includegraphics[width=0.5\textwidth]{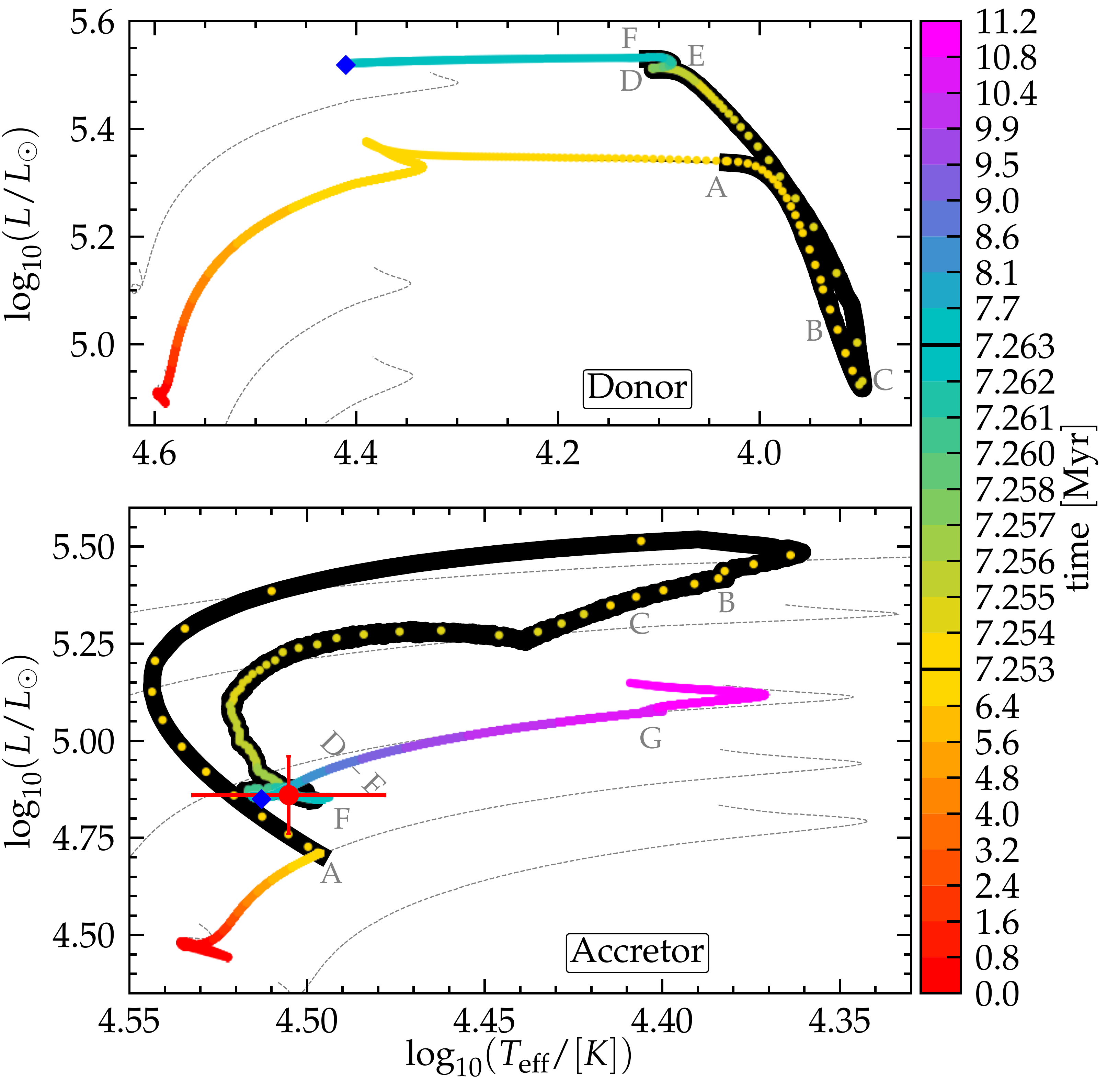}
  \caption{HR diagram for the donor star (top) and accretor star
    (bottom) of a model for an interacting binary star. Each point is
    separated by 50\,years, and the black outline corresponds to the
    RLOF phase. The colors show the age on a non-uniform scale: we use
    smaller time-bins during RLOF
    ($7.25\,\mathrm{Myr}\lesssim t \lesssim 7.26\,\mathrm{Myr}$).  The
    red data point shows the estimated stellar properties of \zoph\ according to
    \citetalias{villamariz:05}. The blue diamonds mark the end of
    the binary run, after which we continue to evolve the accretor as a single
    star from until TAMS, thus the bottom panel
    shows a longer time. We emphasize the different scales on the two
    panels. The thin gray dashed line show the main sequence evolution
    of non-rotating single stars of 15, 17, 20, 25, and 30\,$M_\odot$
    at $Z=0.01$ for comparison.}
  \label{fig:HRD_both}
\end{figure}

We describe here the evolution of a binary system in which the
accretor can broadly reproduce the observed features of \zoph. We
assume initial masses $M_1=25\,M_\odot$, $M_2=17\,M_\odot$, and initial orbital
period $P=100$\,days (corresponding to a separation
$a\simeq314\,R_\odot$) with a metallicity of $Z=0.01$.

\Figref{fig:HRD_both} shows the Hertzsprung-Russell (HR) diagrams of
the donor (top) and accretor star (bottom) -- see \Figref{fig:sp_test} for a single HR
diagram with both stars. Each marker in the figure
corresponds to an elapsed interval of 50\,years.

After $7.24$\,Myr, the most massive star in the system
 evolves off the main sequence and $\sim{}8400$\,years later, at
point A in \Figref{fig:HRD_both}, it overfills its Roche lobe and starts to donate mass. This
results in a stable case B RLOF on a thermal timescale from point A to
F (black outline of the curves).
We refer to \cite{gotberg:17, klencki:20, laplace:21,
blagorodnova:21} and references therein for a detailed description
of the evolution of massive donor stars in binaries.  Although our
models are more massive, the qualitative behavior of the donor star
is similar. Minor differences might arise because of mixing above
and in the H-burning shell \citep[e.g.,][]{schootemeijer:19,
klencki:21}, and its interplay with the mass transfer.

At the onset of RLOF (point A in \Figref{fig:HRD_both}), the accretor
star is still on the main sequence with
$T_\mathrm{eff}\simeq10^{4.5}$\,K and its central mass fraction of
hydrogen is $X(^1\mathrm{H})\simeq 0.42$. The accretion of material
drives the star out of thermal equilibrium and it quickly becomes
over-luminous to radiate away the excess internal energy. The accretor
reaches
$L\simeq10^{5.5}\,L_\odot\gg L_\mathrm{nuc}\simeq 10^{5.1}\,L_\odot$,
with $L_\mathrm{nuc}$ the total energy released per unit time by
nuclear burning (integrated throughout the star). We note that because
of the rejuvenation (see \Secref{sec:mixing}), $L_\mathrm{nuc}$ also
increases from its pre-RLOF value. The
duration of this phase is only $\lesssim 2\times 10^3$\,years,
corresponding roughly to the thermal timescale of the outer envelope
of the accretor. During the same phase, the radius of the
accretor increases dramatically from $\sim{}7.5\,R_\odot$ to
$\sim{}35\,R_\odot$.

At point B -- close to the lowest $T_\mathrm{eff}$ --
the accretor reaches critical rotation, which briefly decreases the
efficiency of mass accretion. With lower mass accretion rate, the star can contract and heat up, causing the track to change direction. The contraction also increases
$\omega_\mathrm{crit}$, allowing for further accretion to resume.

At $T_\mathrm{eff}\simeq 10^{4.42}$\,K, slightly after point C, the
inner layers of the donor's envelope are exposed. These layers were
part of the donor's convective hydrogen-burning core earlier during the main sequence evolution, before the core receded in mass
coordinate. Because this material has been exposed to the CNO cycle, the transferred material becomes progressively
more enriched in helium and nitrogen. The difference in composition of the
incoming material affects the opacity and mean molecular weight in the
outer layers of the accretor and causes a kink in its evolutionary
track, around $\log_{10} (T_{\rm eff}/\mathrm{K}) \sim 4.45$. Specifically, material with a high mean molecular weight $\mu$
is placed on top of the comparatively low-$\mu$ envelope of
the accretor.
Because of the increasing gradient in mean molecular weight,
thermohaline mixing starts in the outer layers of the accreting star,
and, together with rotational mixing, it progressively dilutes the
surface $^4\mathrm{He}$ and $^{14}\mathrm{N}$ mass fractions (we
discuss further mixing processes and the internal composition of the
accretor in \Secref{sec:mixing}). The numerical treatment of these
mixing processes causes noisy features from point D
to F on the HR diagram of the accretor \citep[e.g.,][]{cantiello:07}.

From D to E the donor star expands again: by point D the surface is He-rich, and partial
recombination of $^4\mathrm{He}$ drives a convective layer which has extremely low mass ($\lesssim 10^{-4}\,M_\odot$) but can expand to large radii\footnote{With previous \texttt{MESA} releases, we found it
  challenging to compute models beyond this phase: the large
  increase in radius impacts significantly the mass transfer rate.}.

Accounting for both wind mass loss and the amount of mass transferred,
at the end of our binary run (blue diamonds in \Figref{fig:HRD_both})
the donor becomes a helium star of $\sim 9.4\,M_\odot$, likely to
contract further. Depending on its wind mass-loss rate, the stripped
donor star's spectrum is expected to show absorption lines, emission lines, or a
mixture of both \citep[e.g.,][]{crowther:07, massey:14, neugent:17,
  gotberg:18, shenar:20}. It's surface H mass fraction is $\lesssim 0.2$ and
this residual H-rich layer could be removed by further wind
mass loss \citep[e.g.,][]{yoon:17, gotberg:17, laplace:20}.

We emphasize that our adopted (standard) choices to model mixing and
rotation are likely to impact the morphology of the accretor's
evolutionary track during RLOF. The entire duration of RLOF from A to
F is only about $10^4$\,years - of the order of the thermal timescale
of the donor star. Moreover, the accretor spends most of this time
close to the final, post-RLOF position (blue diamond in the bottom
panel of \Figref{fig:HRD_both}). Therefore, the direct observation of
a (population of) mass-transferring binary(ies) is unlikely.

\subsection{Mass, velocity, photometry, and age of \zoph\ are naturally
  explained by binary interactions}

At the end of our binary evolution, well after the donor detaches from
its Roche lobe (blue diamonds in \Figref{fig:HRD_both}) the accretor
is a H-rich fast-rotating star of $\sim{}20.1\,M_\odot$. This matches
well with the estimates for \zoph, which although highly uncertain,
typically include $20\,M_\odot$ in their range (e.g.,
\citealt{hoogerwerf:01}, \citetalias{villamariz:05},
\citealt{neuhauser:20}).

The post-RLOF orbital velocity of the
accretor is $v_2\simeq52\,\kms$. In the subsequent evolution, wind
mass loss from both stars will widen the binary and decrease the
orbital velocity of the accretor star.  As a test, we evolved one
binary assuming the \cite{nugis:00} wind mass loss rate for the
stripped donor until its He core depletion. At that point the
accretor star's orbital velocity has decreased to $\sim{}40\,\kms$,
and it is expected to decrease even further during the remaining
evolution.
The precise amount of the orbital widening and decrease of
the accretor's orbital velocity depends on the uncertain stripped
star mass loss rate \citep[e.g.,][]{vink:17, sander:20}. Nevertheless,
the value we obtain for the orbital velocity of the accretor is in broad agreement with estimates of the
observed runaway velocity of \zoph.

In the evolution beyond the blue diamond in \Figref{fig:HRD_both}
(computed as a single star) the accretor settles on a main-sequence
track at higher luminosity compared to the original track because of
the accretion of mass, and its slope is slightly steeper due to the
close-to-critical rotation and the accretion of partially nuclear
processed (He- and N-rich) material.

The effective temperature, bolometric luminosity, and kinematic age of
\zoph\ are also reasonably well reproduced by our model (cf.~\Figref{fig:HRD_both}). After
detachment, the donor star has approximately 0.5 Myr left until
core-collapse, which likely will disrupt the binary system and eject
the accretor star \citep[e.g.,][]{renzo:19walk}. As indicated by the colors
of the track in \Figref{fig:HRD_both}, our accretor model spends
around 2\,Myr within the error bars for \zoph\ estimated by
\citetalias{villamariz:05}. This means that after the companion explodes,
the accretor star will look similar to \zoph\ for around 1.5 Myr, in
good agreement with the estimated time  of $1.78\pm0.21$ Myr since \zoph\ was unbound from its companion
\citep{neuhauser:20}.

According to our model, the
present-day age of \zoph\ is $\sim{}7.5-9.5$\,Myr.  The age of the parent
association of \zoph, Upper-Centaurus-Lupus, is relatively uncertain,
with estimates from pre-main sequence isochrone fitting of about
$9\pm1$\,Myr, but an average age of $15\pm3$\,Myr
\citep[][]{pecaut:16}.  Given the sensitivity of our model to
mixing processes and rejuvenation (see also \Secref{sec:mixing}), the large age
scatter of the region, and the possible systematics in age
measurements, we consider our model broadly compatible with the
existing constraints.

We discuss in detail the mass and mass-transfer evolution in
\Secref{sec:MT}, the internal and surface rotation in
\Secref{sec:rot}, and the chemical composition in
\Secref{sec:mixing}. As a summary, the accretor in our binary starts
with 17$\,M_\odot$ and accretes about 3.4\Msun\ material during mass
transfer (out of $\sim{}10.6$\Msun\ total transferred). This causes
rejuvenation: our accretor reaches $T_\mathrm{eff}=32\,000$\,K (the
$T_\mathrm{eff}$ of \zoph\ estimated by \citetalias{villamariz:05}) at
$\sim{}7.3$\,Myrs while a single initially 20\,$M_\odot$ star would reach
such temperature at $\sim$ 6.5\,Myr. Moreover, our accretor reaches
TAMS after 11.2\,Myr, similar to the lifetime of a non-rotating single
star of 17\,$M_\odot$, which is $\sim$11.1\,Myr with our setup. Initially
20\,$M_\odot$ rotating models have a main-sequence lifetime of
$\sim$9.2-9.6\,Myr (longer for higher initial rotation rates). After
mass the mass transfer, the accretor star spins rapidly (see
\Secref{sec:surf_rot}), and its surface composition is determined by
the accretion of material from the donor star, which is progressively mixed
inwards in the accretor (see \Secref{sec:surf_comp}). We find that the rotation and
surface composition of \zoph\ are more easily explained by accretor
models than single rotating stars (e.g., \citetalias{villamariz:05}).

\section{Mass and mass-transfer rate evolution}
\label{sec:MT}

\begin{figure}[tp]
  \includegraphics[width=0.47\textwidth]{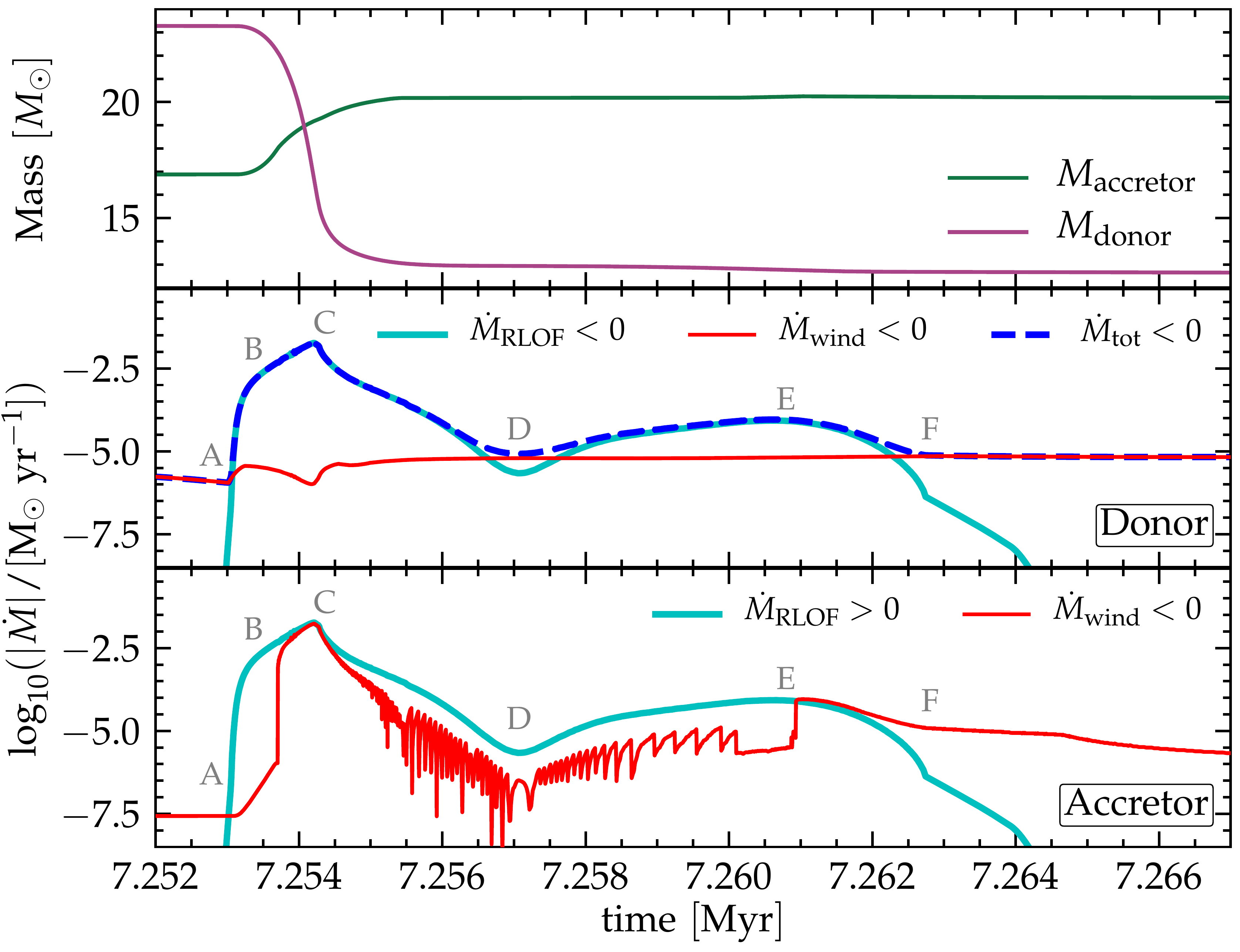}
  \caption{The top panel shows the total mass of each star as a
    function of time during RLOF. The middle and bottom panel show the
    contributions to the rate of mass change for the donor and
    accretor star, respectively. The cyan solid lines show the mass
    transfer rate due to RLOF, the red lines show the wind mass loss rates, which is mechanically enhanced for the accretor during mass transfer. In the middle panel, the dashed
    blue lines shows their sum, corresponding to the total mass loss
    rate of the donor. During RLOF the accretor reaches critical
    rotation, which leads to numerical oscillations in the
    rotationally-enhanced wind mass loss.}
  \label{fig:MT}
\end{figure}

\Figref{fig:MT} shows the mass evolution (top panel) and the rate of
mass change (middle and bottom panels) for each individual star during
the mass transfer phase. The donor star (middle panel) loses mass via
RLOF ($\dot M_\mathrm{RLOF}<0$, cyan line) and wind
($\dot M_\mathrm{wind}<0$, thin red line). The dashed blue line shows
the sum of these two negative terms and represents the total rate of
mass change of the donor. Conversely, the accretor (bottom panel)
gains mass via RLOF (i.e., $\dot M_\mathrm{RLOF}>0$ from the
accretor's point of view), but still loses mass to the wind
($\dot M_\mathrm{wind}<0$).
At peak (point C), the mass transfer rate
reaches above $10^{-2}\,M_\odot\ \mathrm{yr^{-1}}$ and taps
into the optically thick matter of the donor (i.e., the donor's Roche
radius becomes smaller than its photosphere $R_\mathrm{RL,1}<R_1$).

Between point A and B, the mass transfer rate equals the
mass accretion rate (bottom panel of \Figref{fig:MT}), that is,
initially the accretion is (by construction) fully conservative. The
bulk of the accreted mass -- $\sim{}2\,M_\odot$ out of
$\sim{}3.1\,M_\odot$ -- is accreted during this phase, which
lasts about $\sim{}2\times10^3$\,years. As the mass and surface
rotation rate of the accretor increase, the adopted
enhancement of the wind due to rotation progressively increases the
mass-loss rate by $\sim$5 orders of magnitude. This
mechanically-enhanced wind controls the accretion efficiency, and at
$\sim{}7.254$\,Myr (from B to C, where the red solid line and the cyan
line overlap) the mass transfer becomes briefly non-conservative. In
our setup, the majority of the mass that is transferred during this phase is
ejected as a fast wind from the accretor, carrying the same specific
angular momentum as the orbit of the accretor star. The decreased
accretion efficiency allows the accretor to contract. In the remaining evolution from C to F,
the interplay between the wind mass loss rate, the spin-up due to
accretion and the spin-down due to inward transport of angular
momentum (see \Secref{sec:rot}) cause numerical oscillations in the wind
mass loss rate. Nevertheless, for most of the evolution, accretion
still occurs, albeit not-fully conservatively. This allows for
CNO-processed material from the donor to reach the surface of the
accretor during late stages of mass transfer.

The minimum of the mass transfer rate in point D corresponds to a
brief phase of contraction (see also \Figref{fig:HRD_both}). However,
from D-E, the donor star expands again.  This is due to the
partial recombination of the now He-rich outer layers, which causes a
transient surface convection layer. The convective layers expand,
leading to an increase in the mass-transfer rate, despite at this
point 
the binary 
is widening. During this phase, the mass transfer becomes highly
non-conservative again (in the bottom panel the wind and the mass
accretion rate nearly cancel each other again at $\sim7.261$\,Myr),
until the donor completely detaches from its Roche lobe at point F.

\section{Rotation and angular momentum transport in the accretor}
\label{sec:rot}

\subsection{Surface rotation}
\label{sec:surf_rot}

\begin{figure}[bp]
  \includegraphics[width=0.5\textwidth]{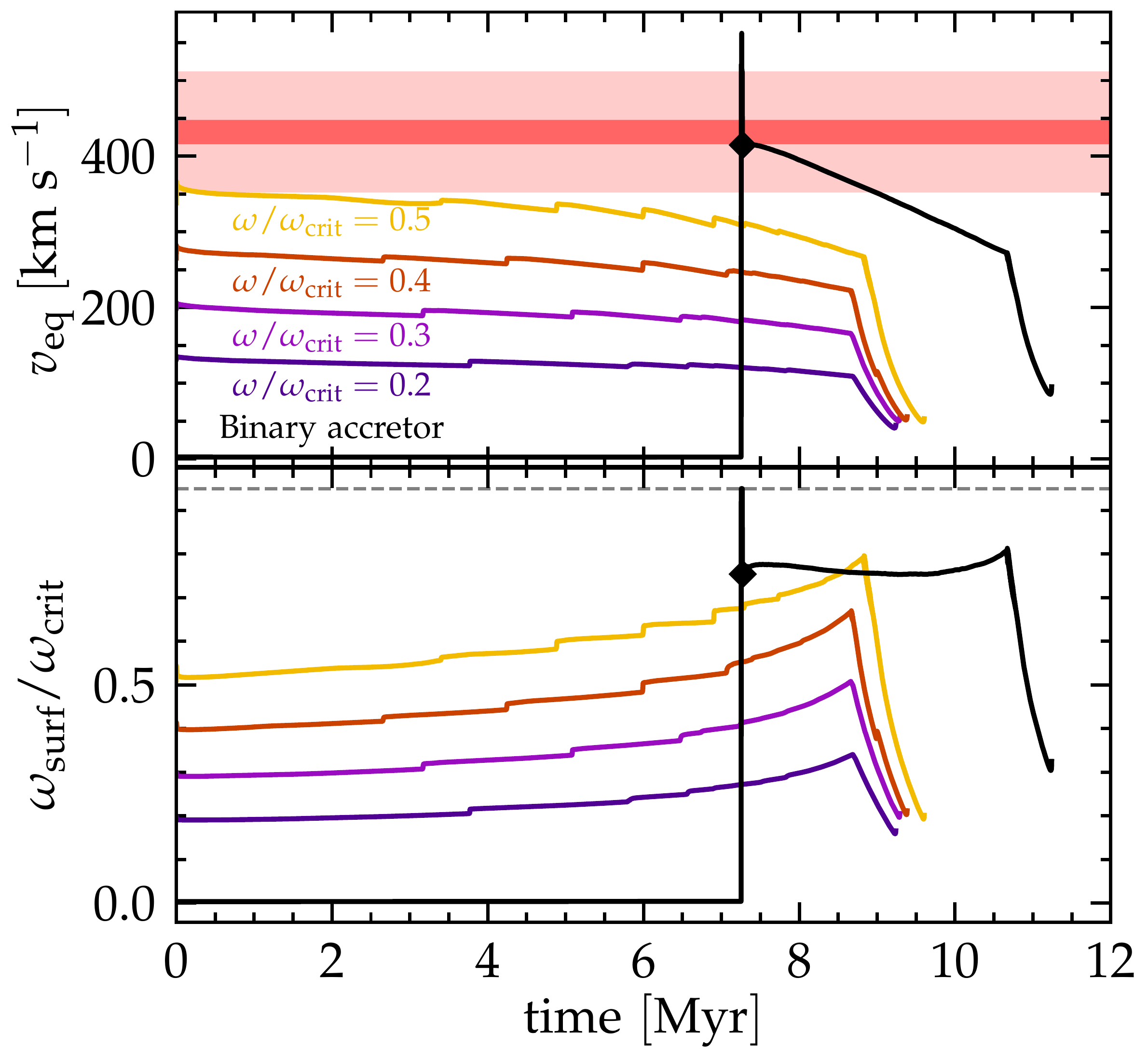}
  \caption{Equatorial surface rotational velocity (top panel) and
    $\omega/\omega_\mathrm{crit}$ (bottom panel) for the accretor
    model (black) and single rotating $20\,M_\odot$ stars (colored
    lines). The red bands in the top panel correspond to the
    $v\sin(i)$ observed for \zoph\ (see text). At $\sim{}7.25$\,Myr
    the mass transfer quickly spins up the accretor to critical
    rotation: the dashed horizontal line in the bottom panel shows the
    upper-limit we impose. By the time the donor detaches from the
    RLOF the accretor is still spinning at $\sim{}400\,\kms$. From
    this point (diamonds) onwards, we continue the evolution as a
    single star, and the accretor spins down because of wind mass
    loss.  Note however that we use the wind mass-loss rate from
    \cite{vink:00, vink:01}, which is suggested to be overestimated  $\sim$2 orders of magnitude \citep{marcolino:09}.}
  \label{fig:surf_rot}
\end{figure}

Rapid rotation is one of the main properties expected as a result of
mass accretion \citep{packet:81}.  One of the main
distinguishing features of \zoph\ is its extremely high surface
rotation rate. The black line in the top panel of
\Figref{fig:surf_rot} shows the evolution of the surface equatorial
rotational velocity $v_\mathrm{eq}$\footnote{More precisely, $v_\mathrm{eq}$ is a
mass-weighted average of the equatorial rotation velocity for the layers with
optical depth $\tau\leq 100$.} for our accretor model, not
including any projection effect. In the top panel, the dark horizontal red band shows the velocity  $432\pm16\,\kms$, which is the $v \sin (i)$ measured for \zoph\ by \cite{zehe:18}. The lighter red band shows a range of 5 times their error bar, which
roughly encloses the majority of the estimated $v\sin(i)$ for \zoph\
in the literature $350\,\kms\lesssim v\sin(i)\lesssim 600\,\kms$
(e.g., \citealt{gordon:18} and \citealt{walker:79}, respectively).
For comparison, the colored solid lines show also $v_\mathrm{eq}$ for
models of rotating $20\,M_\odot$ single stars with birth spins of
$\omega/\omega_\mathrm{crit} = 0.2$, 0.3, 0.4 and 0.5. We note that
the ``steps'' in these lines are numerical artifacts which do not
impact our results. The bottom panel of \Figref{fig:surf_rot}
shows the ratio of the surface rotational frequency
$\omega_\mathrm{surf}$ to the critical rotational frequency
$\omega_\mathrm{crit}$.

The initial binary is wide enough that assuming tidal synchronization
at ZAMS implies a very low $v_\mathrm{eq}\lesssim 3\,\kms$. At 7.25\,Myr, mass transfer rapidly spins
up the accretor to critical rotation,
$v_{\rm crit}\sim{}520\,\kms$. This corresponds to
$\omega/\omega_\mathrm{crit}\simeq 0.95$ (dashed horizontal line in
the bottom panel of \Figref{fig:surf_rot}), which is the upper-limit
we impose in our models (see \Secref{sec:methods}).

The star remains fast rotating during the mass transfer phase, and throughout
the remaining evolution in a binary, which ends at the black diamond in
\Figref{fig:surf_rot} (corresponding to the blue diamonds in
\Figref{fig:HRD_both}). Afterwards, the star spins
down progressively because of wind mass loss, and within $\sim$2\,Myr its
equatorial surface rotational velocity drops below $\sim{}350\,\kms$.

Both the single star models and our accretor (after being spun up)
evolve to higher $\omega/\omega_\mathrm{crit}$ because of the increase
in stellar radii and corresponding decrease in $\omega_\mathrm{crit}$
\citep[e.g.,][]{langer:98, zhao:20}. However, our accretor model
remains at a higher
$\omega_\mathrm{surf}/\omega_\mathrm{crit}\simeq 0.75$ for a
significantly longer time: the chance of observing a single rotating
star at very high $\omega_\mathrm{surf}/\omega_\mathrm{crit}$ is lower
than for an accretor from a massive binary system. Moreover,
the $\omega_\mathrm{surf}/\omega_\mathrm{crit}$ we find is in good
agreement with the observationally constrained values for typical Oe
and Be stars \citep[see][for a review]{rivinius:13}.

Close to the end of the main sequence, the increase in wind mass loss
rate as the stars cross the bistability jump \citep[due to iron recombination
at $T_\mathrm{eff}\simeq25\,000\,\mathrm{K}$, e.g.,][]{vink:00}
strengthens the surface spin-down. This effect is also seen in the
late main-sequence evolution of single stars rotating rapidly at
birth.

\Figref{fig:surf_rot} shows that our model retains a significant
surface rotation for a long period of time, comparable to the
kinematic age of \zoph. Since the spin up of the accretor happens
roughly half-way through its main sequence, the star is much faster
rotating than single stars of the same mass, but which were initialized as
fast rotators at ZAMS. Although \Figref{fig:surf_rot} does not account
for the projection angle, $i$, \cite{zehe:18} argued for
$i\geq56$\,degrees, corresponding to an upward shift of the red band
in \Figref{fig:surf_rot} of $\lesssim 20\%$. This shift impacts the
comparison of \zoph\ to our accretor model and to single star models
in the same way.

We emphasize that our model is computed using the \cite{vink:00,
  vink:01} wind mass-loss rate with full efficiency throughout its
evolution. This is two orders of magnitude higher than the wind mass
loss rate reported by \cite{marcolino:09} (weak wind problem,
however, see also \citealt{lucy:12, lagae:21}). While this may impact
the evolution of the binary even before RLOF, it presumably increases the
spin-down rate of our model compared to the observations. We expect
that an accreting star modeled with lower wind-mass loss rate
after mass transfer would retain an even higher surface rotation for longer (see
also \Secref{sec:single_star_uncertainties}).

\subsection{Internal rotation -- comparison to  single stars}
\label{sec:rot_comparison_single}

\begin{figure*}[tbp]
  \centering
  \includegraphics[width=\textwidth]{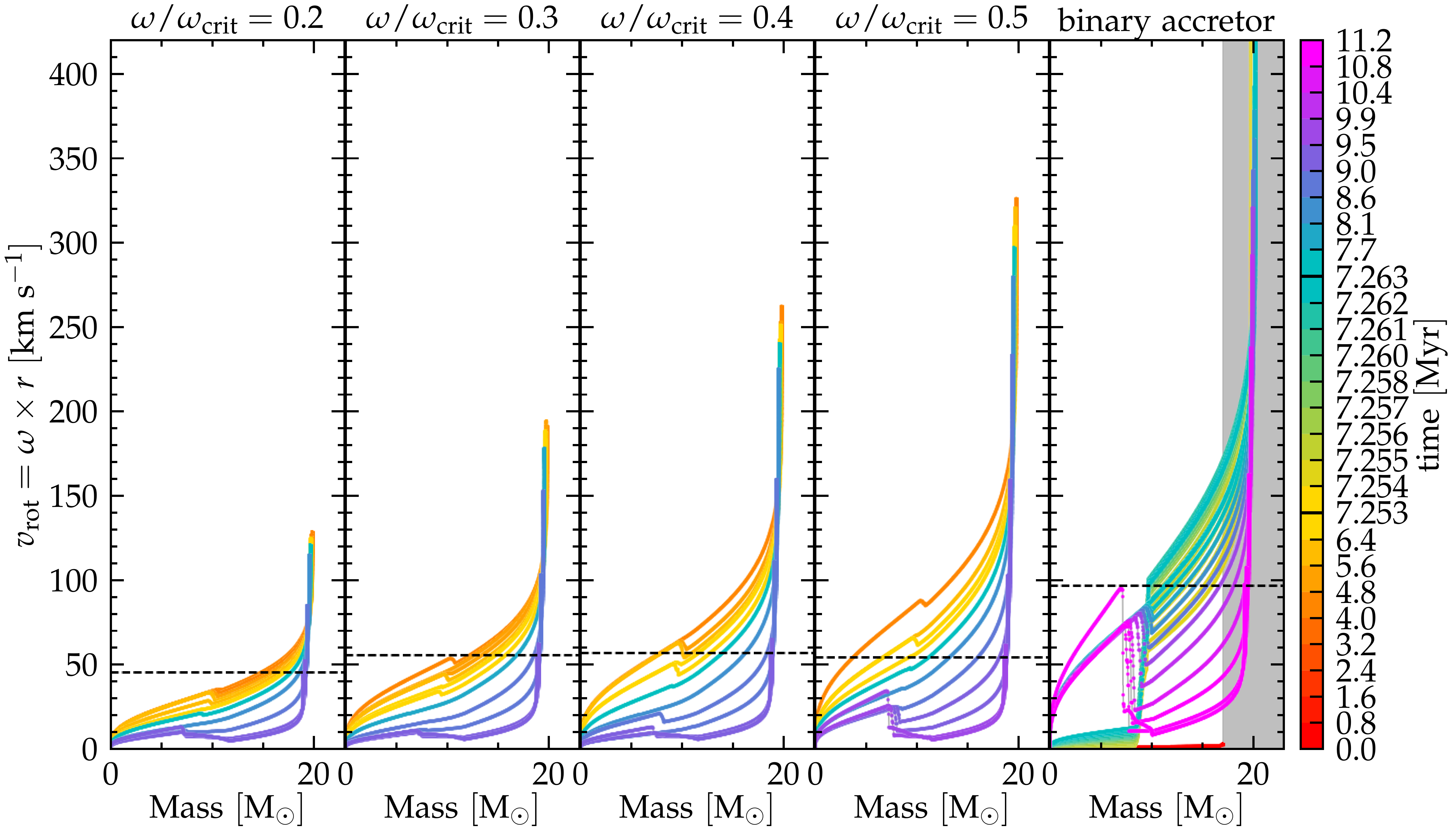}
  \includegraphics[width=\textwidth]{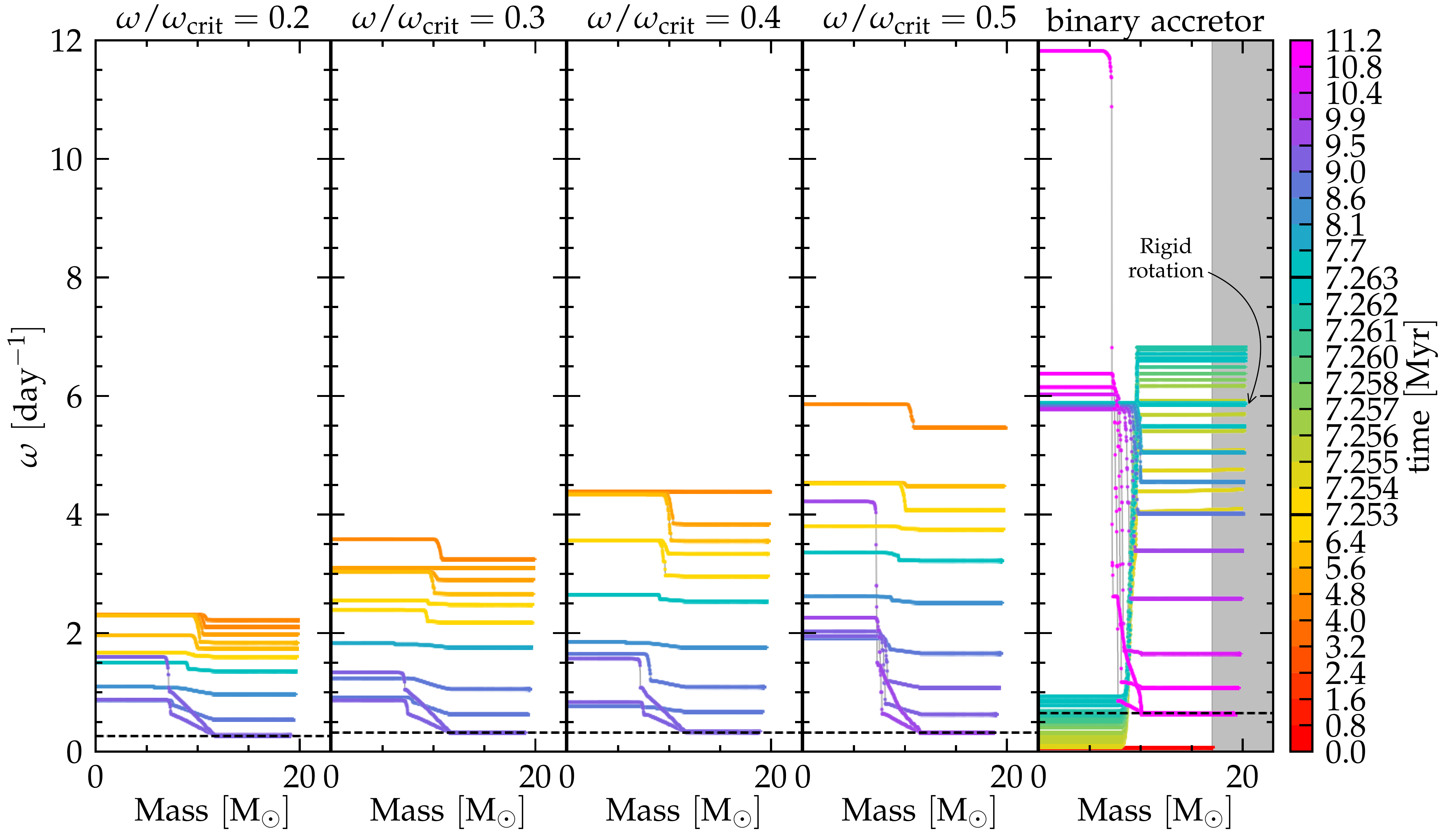}
  \caption{Internal rotational velocity (top) and frequency
    (bottom) profile and for $20\,M_\odot$ single star models with
    increasing $\omega/\omega_\mathrm{crit}$ at birth (first four
    panels), and for the accretor of our fiducial binary. As in
    \Figref{fig:HRD_both}, the colorbar shows stellar age and is non-uniform. In the top
    (bottom) panel the thin dashed black lines mark the surface
    rotation rate (surface rotation frequency) at the end of the main sequence (TAMS). In the rightmost
    panels, the gray areas indicate mass accreted during mass transfer. The
    pink lines (structure at TAMS) in the last panel show that the core of the
    accretor is rotating almost as fast as its surface despite its
    much smaller radius, and both are rotating faster than the surface of
    the single star models.}
  \label{fig:struct_rot}
\end{figure*}

Our model suggests that the internal rotational profile of accretor
stars evolve differently compared to those of rotating single stars.
To illustrate the angular momentum transport in our accretor star, it
is again helpful to compare with single rotating $20\,M_\odot$
models.


The top row of \Figref{fig:struct_rot} shows the internal rotational
velocity, $v_\mathrm{rot}=\omega\times r$, while the bottom row shows
the angular frequency profile $\omega$ as a function of mass
coordinate. The first four panels in each row show single rotating
stars with increasing initial $\omega/\omega_\mathrm{crit}$ and the last
panels show our accretor model. The gray area in the rightmost
panels highlights matter accreted during RLOF, while the colors of the
lines indicate the age of the star for each profile that is shown.

The thin dashed lines in each panel of the top row of
\Figref{fig:struct_rot} mark the TAMS surface rotation rates: all our
single star models reach a TAMS surface
$v_\mathrm{eq}\simeq50-60\,\kms$ (see \Figref{fig:surf_rot}). Initially faster rotating models
spin down more in their outer layers and have slightly longer main
sequence lifetimes (because of rotational mixing increasing the
available fuel). As the
core contracts and spins up, the single star profiles show a slight
progressive development of a core-envelope interface.

Conversely, the entire interior of the accretor has a negligible
rotational velocity until RLOF (starting at
$\sim{}7.25$\,Myr). Because of binary mass transfer, the accretor is
spun up from the surface inwards (cf.\ initially rigid rotation assumed
for single stars, \citealt{maeder:00}), late in its evolution (cf.\
at ZAMS for single stars), and it
reaches critical rotation $\omega/\omega_\mathrm{crit}\simeq1$. In our
model, inward transport of angular momentum creates a $v_\mathrm{rot}$
profile monotonically decreasing from the surface to the center. After
the end of mass transfer, roughly at $7.27$\,Myr, the accretor
achieves rigid and close to critical rotation (flat profiles in the
last panel on the bottom row of \Figref{fig:struct_rot},
$\omega \sim 6$ day$^{-1}$). At the end of our binary run the
accretor is still rigidly rotating, which persists\footnote{To calculate
  the duration, we consider rotation to be rigid if the difference
  between the minimum and maximum frequency throughout the star is
  $\Delta \omega \lesssim 10^{-2}\,\mathrm{days^{-1}}$.} for a total
duration of $\sim10^{4}$\,years. Afterwards, the accretor's envelope
spins down because of winds and its evolutionary expansion.  By the
end of the accretor's main-sequence, the surface still spins with
$v_\mathrm{rot}\simeq100\,\kms$, which is approximately twice as fast
as the single star models.

Conversely, in the remaining evolution, the core contracts (decrease
in $r$). The weak core-envelope angular momentum coupling provided by
the Spruit-Tayler dynamo leads to an approximately constant total
angular momentum in the core, therefore, as it contracts and decreases
its moment of inertia, its rotation rate $\omega$ increases.  At the end of the main sequence, the outer edge of the
core of the accretor star has a similar rotational velocity as the
surface, and much larger than for the single star models.
Consequently, the TAMS core-envelope interface
for the accretor is much more prominent than for in single rotating
stars. It might be possible to distinguish accretors from
initially single stars by using asteroseismology to measure the core
rotation rate \citep[e.g.,][]{cantiello:14}.
However, this requires
the detection of mixed modes which is presently challenging for
massive stars.

Moreover, the higher core spin of accretors may have important
implications for their future explosions \citep[e.g.,][]{macfadyen:99,
  cantiello:07}, the spin of the resulting compact objects, and the
analysis of gravitational-wave events \citep[for systems other than
the progenitor of \zoph, which remain bound and can evolve into
gravitational wave mergers, e.g.,][]{zaldarriaga:18,
  qin:18, callister:21}.

\section{Mixing and composition of the accretor}
\label{sec:mixing}

\begin{figure}[htbp]
  \includegraphics[width=0.47\textwidth]{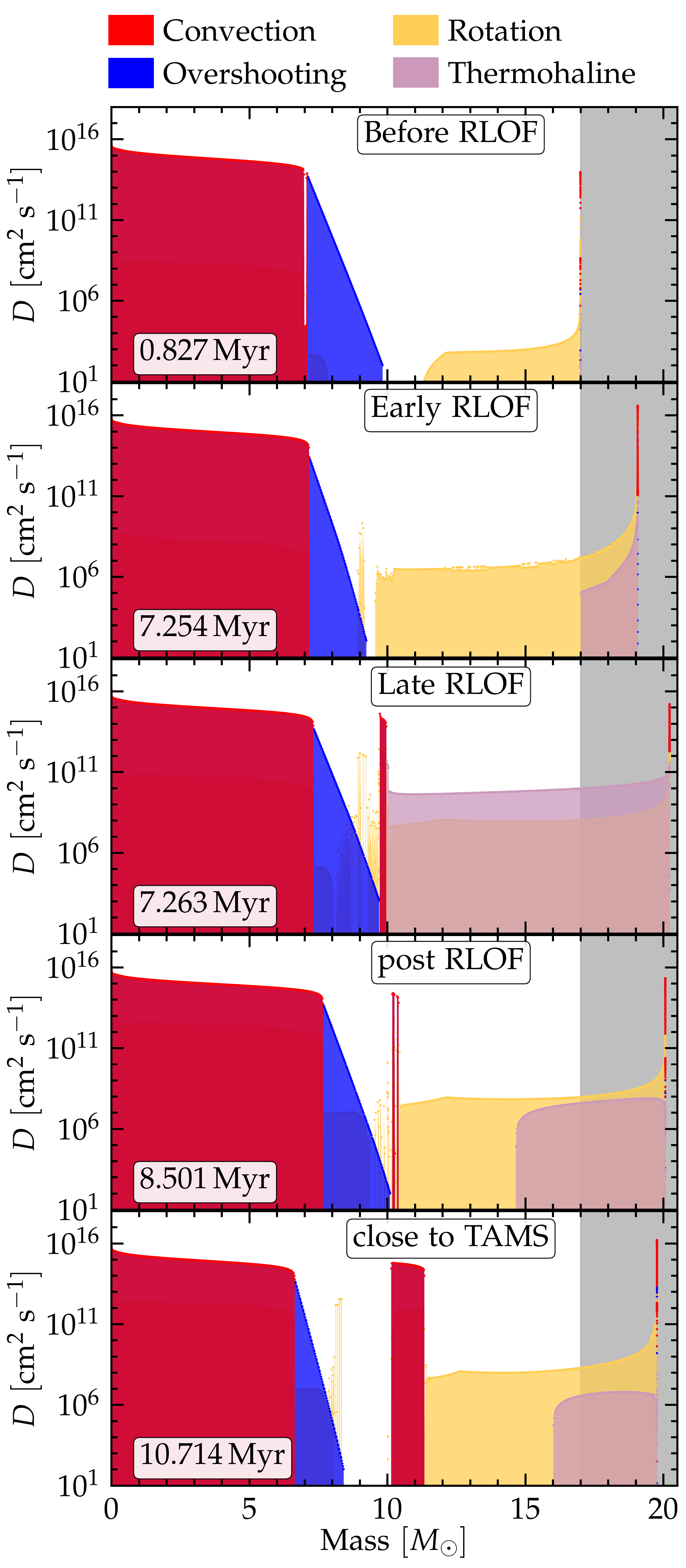}
  \caption{Mixing diffusion coefficients in the accretor star as a
    function of mass (the center corresponds to abscissa 0 and the
    surface to the maximum abscissa for which a diffusion coefficient
    is plotted). The gray area on the right highlights accreted
    material. From top to bottom, each panel shows a profile during
    the main sequence (before point A in \Figref{fig:HRD_both}), early
    during RLOF (slightly after point C), late during RLOF (close to
    point F), after detachment (within the \zoph\ errorbar on
    \Figref{fig:HRD_both}), and after RLOF (point G). A movie of the
    entire evolution is \referee{included in \Figref{fig:mix_movie}}.}
  \label{fig:D_mix}
\end{figure}

Rotational mixing and thermohaline mixing induced by the accretion of
CNO-processed material affect significantly the composition profile of
our accretor star. They both act primarily in the envelope, starting
from the surface and growing inwards. \texttt{MESA} treats mixing
using a diffusion approximation \citep{paxton:11}, and to illustrate
the dominant processes we show in \Figref{fig:D_mix} the diffusion
coefficients for the mass fractions of elements as a function of
Lagrangian mass coordinate at selected times.  In each panel of
\Figref{fig:D_mix}, the gray background highlights mass coordinates
exceeding the initial mass of the star, reached because of accretion.
The remaining colors show convection (red), overshooting (blue),
rotational mixing (yellow), and thermohaline mixing (pink). Rotational
mixing includes all the rotational instabilities that we consider --
meridional currents, secular and dynamical shear instabilities, and
GSF instability. However, throughout the evolution the meridional
currents (Eddington-Sweet circulations) dominate. The only exception
is at the interface between core and envelope (i.e., at the outer edge
of the overshooting region), where significant dynamical shear and GSF
mixing occur. For clarity, we do not show semiconvection which never
dominates the mixing in our accretor model.

The top panel shows the typical structure of a main sequence massive
star: the convective core initially reaching $\sim{}7\,M_\odot$ with
the overshooting extension to $\sim{}9\,M_\odot$. The slow initial
rotation causes a very weak meridional circulation in the envelope. Meridional
circulations are also present in the core throughout the evolution,
but with a diffusivity more than nine orders of magnitude lower than
core convection. A small sub-surface convective zone is also
appreciable at the very surface (see, e.g., \citealt{cantiello:21}).

In the second panel from the top (slightly after point C), the star has already accreted
$\sim{}2\,M_\odot$ (extending in the gray region), including some
CNO-processed material from the layers close to the core of the donor star, and its
surface is already spun up to $\sim{}330\,\kms$. Thermohaline mixing
has started in the newly accreted layers, but it is subdominant
compared to rotational mixing due to meridional circulations in the
envelope. Angular momentum transport (by the Spruit-Tayler dynamo) has
already imparted some rotation to the inner layers of the
envelope. This leads to the disconnected spike in rotational mixing at
the outer edge of the core (at mass coordinate $\sim{}9\Msun$),
dominated by dynamical shears.  We note that between the top panel and
the onset of RLOF, the convective core recedes in mass coordinate,
but, by the time shown in the second panel, the accretion of mass has
caused the core to grow back to its initial size.

In the third panel, the star has finished its mass accretion and is
expelling the last transferred material (close to point
F). Thermohaline mixing takes over the dominant role for chemical
mixing in the envelope (although meridional circulations persist
behind it, with a mixing coefficient two orders of magnitude
smaller). \referee{The chemical profile in the innermost layers
  reached by thermohaline mixing is set by the mixing at the
  core-envelope at boundary at $\sim9-10\Msun$, dominated by GSF
  instability and shears up to this point, although now a thin
  convective layer is appearing at this location.}  The core is still
growing in mass, meaning that rejuvenation is still on-going.  Because
we adopt an exponentially decreasing overshooting, the diffusion
coefficient at the outer edge of the overshooting region is small, and
therefore is the mixing between the core and the off-center convective
region weak: this off-center convective region does not inject a
significant amount of H-rich material in the core and is not
participating significantly to the rejuvenation. We do not include
over/undershooting for off-center convective layers, which could
increase the coupling between these layers.

After detachment (fourth panel in \Figref{fig:D_mix}, roughly
corresponding to \zoph's structure today), thermohaline mixing is
progressively shutting down from inside out as mass accretion has
stopped, while meridional circulations remain active because of the
rapid rotation. At this stage, the accreted material is well-mixed
into the envelope of the accretor and there is no composition gradient
in the outermost layers of the envelope.

Finally, the last panel displaying the interior mixing shortly before
TAMS shows that, as the post-RLOF evolution proceeds, the previously
thin, off-center convective layer grows thicker, including almost
1.5\Msun\ of material between the mass coordinates $10-11.5\Msun$. In
addition, the hydrogen-burning core has receded, leaving a
$\sim 1.5\Msun$ thick layer that is unaffected by mixing, meaning
that the core and the convective layer have now completely disconnected.

\begin{figure}[tbp]
  \includegraphics[width=0.47\textwidth]{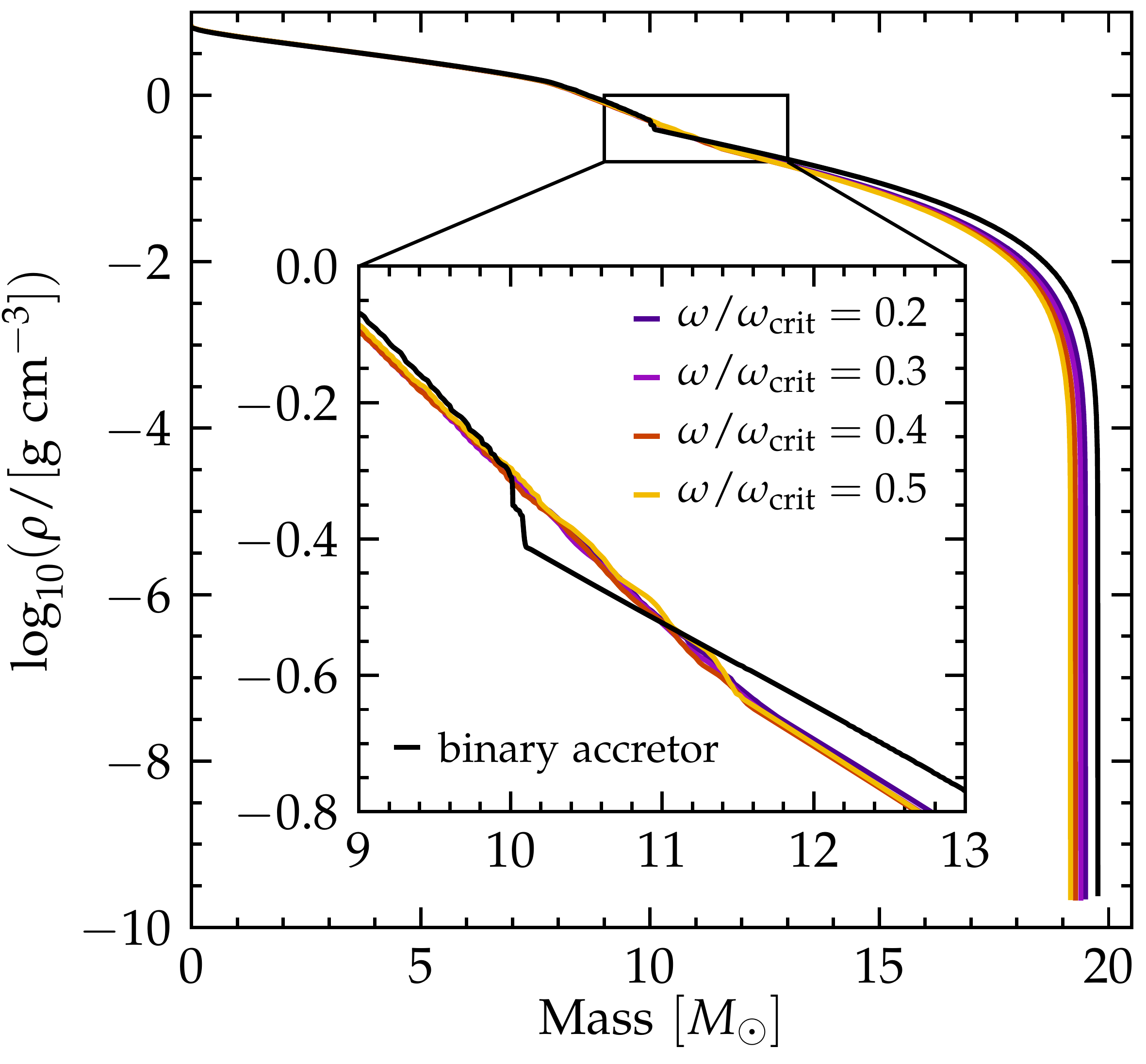}
  \caption{Comparison of the density profiles of the accretor model
    (black) and single $20\,M_\odot$ rotating stars (colors). The
    models are compared when they reach the same central H mass
    fraction of $X_c=0.085$ (point G in \Figref{fig:HRD_both} for the
    accretor). The inset magnifies the region above the core, where
    the outcome of common envelope events is decided. A convective layer has altered the density profile of the accretor.}
  \label{fig:rho}
\end{figure}

A convective layer above the H-burning core
during main-sequence evolution is uncommon for 20\Msun\ single star
models at Z=0.01 (see, e.g.,  \citealt{schootemeijer:19} for lower metallicity
and higher mass models showing off-center convective layers on the
main-sequence). Apart from impacting the composition and density
profile of the star, it is possible that it would also affect the
future evolution of the star.

To illustrate the effect of the off-center convective layer that
develops in the accretor model, \Figref{fig:rho} shows a comparison of
the density profiles between the accretor model (point G in
\Figref{fig:HRD_both}) and our four initially $20\,M_\odot$ single
rotating stars when they reach the same central H mass fraction
$X_c=0.085$ (close to TAMS). Convection significantly alters the
density profile above the core of the accretor compared to the single
rotating stars. The sharper inner density drop and shallower density
profile of accretors may be possible to explore with asteroseismology:
\zoph\ itself has been observed to show non-radial pulsations
\citep{walker:05}, which may be involved in the transient appearance
of emission lines and a decretion disk. If compared to a star of
comparable mass which evolved as single, the pulsations could shed
light on the structural differences between single stars and binary
products.

Moreover, for systems other than \zoph's progenitor, if the binary
remains bound after the explosion of the donor and the evolutionary
expansion of the accretor leads to the development of a common
envelope \citep[e.g.,][]{paczynski:76}, the layer above the He core is
crucial for determine the success or failure of the common envelope
ejection \citep[e.g.,][]{klencki:21}. This might be crucial for our
understanding of gravitational wave progenitors. Common envelope
simulations have so far neglected the impact of previous mass
accretion phase(s) on the density structure of the stars initiating
the dynamically unstable mass transfer.

\begin{figure*}[tbp]
  \centering
  \includegraphics[width=\textwidth]{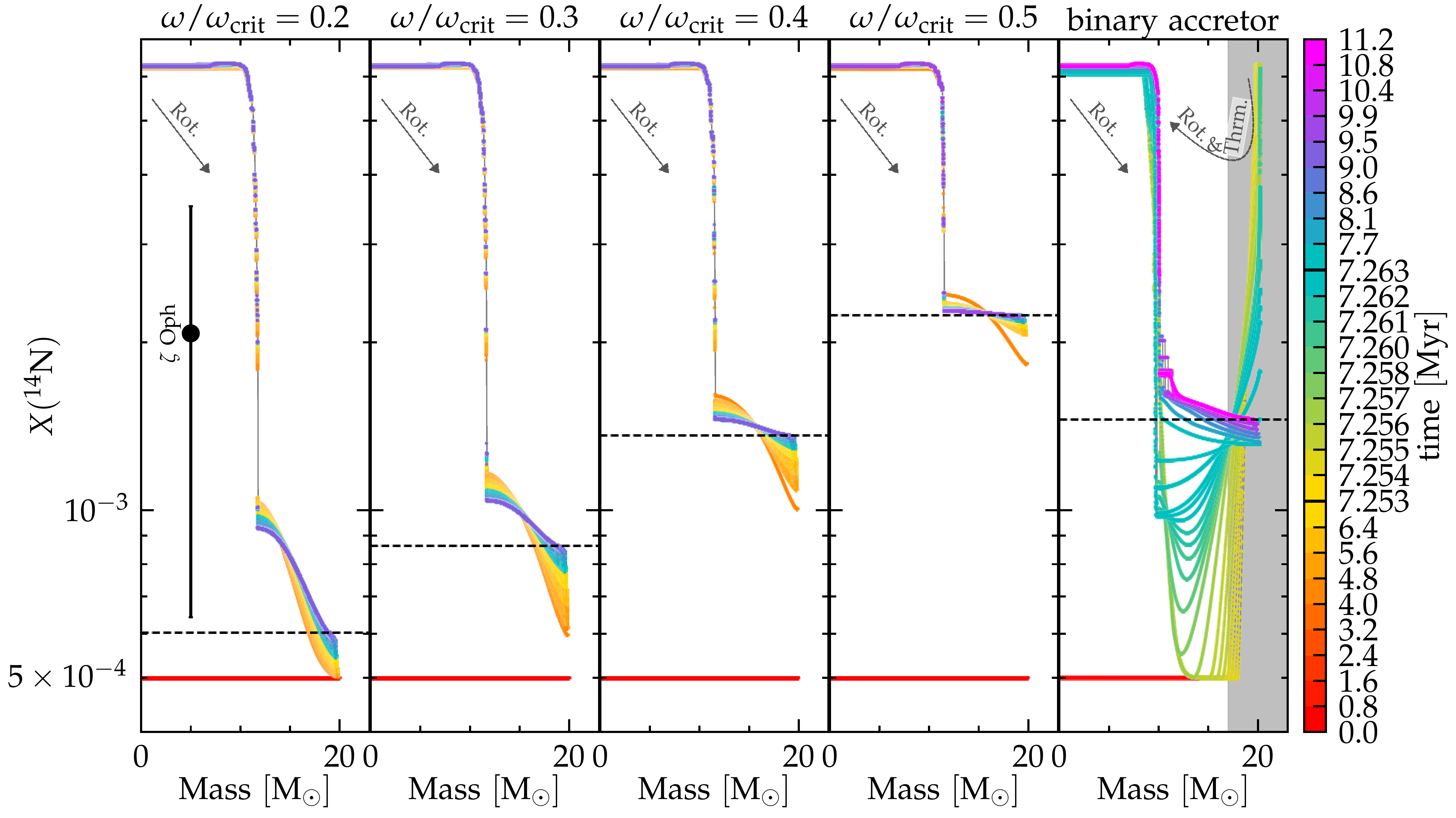} 
  \caption{$^{14}\mathrm{N}$ mass fraction as a function of mass
    coordinate for $20\,M_\odot$ single star models with increasing
    $\omega/\omega_\mathrm{crit}$ at birth (first four panels), and
    for the accretor of our fiducial binary (last panel). The colors
    indicate stellar ages. In each panel, the red flat line marks the
    primordial value for $Z=0.01$, the thin black dashed line marks
    the surface value at TAMS. In the last panel, the gray area
    highlights mass accreted during RLOF. The black errorbar in the
    first panel shows the surface $^{14}\mathrm{N}$ mass fraction of
    \zoph\ estimated by \citetalias{villamariz:05} assuming the
    surface mass fraction of H from \Tabref{tab:surf_prop}. The
    abundance of $^{14}\mathrm{N}$ alone is not strongly constraining.
  }
  \label{fig:n14}
\end{figure*}

\subsection{Internal composition profile -- comparison to single stars}
\label{sec:mix_comparison_single}

\referee{In our models of rotating single stars, the surface
  enrichment in CNO processed material occurs because of outward
  mixing due to rotation, in particular meridional circulations. For
  our accretor model, enriched material accreted from the companion
  also significantly contributes.}

\referee{The result is enhanced abundances of $^{4}\mathrm{He}$ and
  $^{14}\mathrm{N}$. During RLOF, these reach maximum values of
  $X(^{4}\mathrm{He})\simeq0.65$ and
  $X(^{14}\mathrm{N})\simeq6\times10^{-3}$. Subsequently, these
  elements are diluted by inward mixing and partially ejected by the
  accretor's wind.}  After RLOF, the surface $^4\mathrm{He}$ mass
fraction of the accretor has increased by $\sim{}20\,\%$, from
\referee{its primordial value of} 0.26 up to 0.31, while the surface
$^{14}\mathrm{N}$ mass fraction has increased by a factor of 2.5, from
$5\times 10^{-4}$ up to $1.346 \times 10^{-3}$ (see also
\Tabref{tab:surf_prop}). \referee{In the remaining main sequence
  lifetime, the convective layer discussed above couples the core with
  the inner envelope, and rotational mixing brings at the
  surface some of the CNO-processed material from the accretor's own core.} 

Because of \referee{its} large change, we focus here on the
$^{14}\mathrm{N}$ enrichment, but refer to
\Figref{fig:composition_huge} for the evolution of the mass fractions
of \referee{$^{4}\mathrm{He}$}, $^{12}\mathrm{C}$ and
$^{16}\mathrm{O}$. We also include a movie of the evolution of the
accretor's internal composition profile for \referee{these four} elements
plus $^1\mathrm{H}$ \referee{in \Figref{fig:composition_movie}}.
The first four panels of
\Figref{fig:n14} show the mass fraction of $^{14}\mathrm{N}$ as a
function of mass coordinate along the evolution of four $20\,M_\odot$
stars initialized with $\omega/\omega_\mathrm{crit}=0.2,0.3,0.4,0.5$.
The last panel of \Figref{fig:n14} shows our accretor model.
The first four panels of \Figref{fig:n14} show typical rotational
mixing profiles: $^{14}\mathrm{N}$ almost instantly reaches equilibrium in the core because
of the efficient convective mixing in the CNO-cycle in the H-burning core, and it is then mixed outwards (as indicated by the
arrows in the top left corner of each panel).
The $^{14}\mathrm{N}$ profile is monotonically decreasing with mass
coordinate, and the higher the initial rotation, the higher the
surface $^{14}\mathrm{N}$ mass fraction reached at TAMS, because of the
more efficient rotational mixing.

Conversely, the $^{14}\mathrm{N}$ mass fraction profile of the
accretor is \emph{not} monotonic throughout the evolution.  This is because the
profile is shaped by two main processes: (i) accretion of
CNO-processed material from the donor star, which is mixed inwards by thermohaline mixing and meridional
circulations (as indicated by the top right
arrow in the last panel), and (ii) outward mixing of the CNO-processed
material from the accretor's receding main sequence core, brought
ouwards by rotational mixing -- as in the single stars.

Initially, the tidally synchronized accretor star rotates too slowly
($v_{\rm eq} \lesssim 3\,\kms$) for rotational mixing to be efficient
at mixing material outwards from the core. Therefore, until the onset
of RLOF (roughly after 7.25\,Myr) no appreciable variation of the
envelope $^{14}\mathrm{N}$ mass fraction occurs.  Late during RLOF,
close to point F in 
\Figref{fig:HRD_both}, N-enriched material from layers above the
donor's core piles onto the accretor's surface (inside the gray area
in \Figref{fig:n14}). At this stage, the close-to-critical rotation
drive Eddington-Sweet circulations and the inversion in the mean
molecular weight $\mu$ make thermohaline mixing efficient. These
mechanisms drive inward mixing of the accreted N-rich material, while
rotational mixing bring N-rich material outwards, diluting it in the
envelope (see also \Figref{fig:D_mix}, third panel).

Simultaneously, the mere growth in mass causes the steepening of the
core-temperature gradient and increase in the convective core mass
\citep[rejuvenation, e.g.,][]{schneider:16}, driving some outward
convective mixing of N-rich material.  The evolution of the structure
modified by accretion also causes the formation of an off-center
convective region (cf.~\Figref{fig:D_mix}) which persists at least
until TAMS, when we stop our model. Because the convective turnover
timescale is much shorter than the main-sequence nuclear timescale,
convection almost instantly homogenizes the composition of this region
and produces ``steps'' at the outer edge of the core (slightly outside
mass coordinate 10\,$M_\odot$).

\newpage
\subsection{Comparison to \zoph's composition, radius, and rotation rate}
\label{sec:surf_comp}

\begin{table*}[hbt]
  {
    \centering
    \begin{tabular}{c|c|c|c|c|c|c|c|c}
      \hline\hline
      $M \ [M_\odot]$ & $R\ [R_\odot]$ & $ \omega \ [\mathrm{days^{-1}}]$ & $v_\mathrm{rot} \ [\kms] $ & $X(^{1}\mathrm{H})$ & $X(^{4}\mathrm{He})$ & $X(^{12}\mathrm{C})$ & $X(^{14}\mathrm{N})$ & $X(^{16}\mathrm{O})$ \\
      \hline
      20.1 & 9.8 & 4.575 & 361.4 & 0.678044 & 0.312058 & 0.001339 & 0.001346 & 0.004149 \\
      \hline
    \end{tabular}
    \caption{Surface properties of the accretor close to the
      present-day position of \zoph\ on the HR diagram,
      corresponding to a physical age of $8.50\referee{1}$\,Myr, shortly after the end of RLOF
      (between the blue diamond and the lower estimate of
      \zoph's $T_\mathrm{eff}$ in \Figref{fig:HRD_both}, fourth panel of \Figref{fig:D_mix}).}
    \label{tab:surf_prop}
    }
\end{table*}


\Tabref{tab:surf_prop} summarizes the surface properties of the
accretor star at 8.5\,Myr (fourth panel of \Figref{fig:D_mix}),
roughly corresponding to \zoph's position on the HR diagram
today. Both the mass and radius agree reasonably well with the
estimates from \citetalias{villamariz:05} and previous studies, that
is~$20\,M_\odot$ and $8.3\pm1.5\,R_\odot$, respectively. Our radius of
$9.8\,R_\odot$ is larger by $\sim0.6\,R_\odot$ than the equatorial
radius recently measured by \cite{gordon:18}, and our model has
$T_\mathrm{eff}\simeq31\,300$\,K, on the lower end of the range
considered by \citetalias{villamariz:05}. The equatorial
rotational velocity in excess of $350\,\kms$ is also in the correct
ballpark albeit possibly on the low end.

We report a surface H mass fraction\footnote{We obtain the mass
  fractions of individual elements inverting the definition
  $\varepsilon(X)=12+\log_{10}(N_X/N_H)$, where $N_X$ and $N_H$ are
  the number fractions of species $X$ and H, respectively
  \citep[e.g.,][]{lodders:19}.} lower than initial because of the
accretion of nuclearly processed material, and the surface mass
fraction of the most prominent species $^4\mathrm{He}$,
$^{12}\mathrm{C}$, $^{14}\mathrm{N}$, $^{16}\mathrm{O}$.  Assuming our
surface H mass fraction $X(^1\mathrm{H})$, the corresponding mass
fractions of $^4\mathrm{He}$, $^{12}\mathrm{C}$, $^{14}\mathrm{N}$,
$^{16}\mathrm{O}$ obtained by \citetalias{villamariz:05} are
$0.34^{+0.14}_{-0.05}$, $0.0006\pm0.0004$, $0.002\pm0.001$, and
$0.005\pm0.004$.

In \Figref{fig:n14}, the black errorbar in the first panel shows
\zoph's surface $^{14}\mathrm{N}$ from \citetalias{villamariz:05}
(assuming the surface H mass fraction from our model listed in
\Tabref{tab:surf_prop}). The mass fraction of $^{14}\mathrm{N}$ alone
is not sufficient to distinguish between these models, and already a
moderate $\omega/\omega_\mathrm{crit}\geq0.3$ is sufficient for models
to reach the lower limit of the error bar.

By the accretor's TAMS, rotational mixing (in the
form of Eddington-Sweet circulations) and thermohaline mixing nearly
homogenize the composition of the envelope of our accretor's
model. The surface mass fractions we obtain are sensitive to the
interplay between several poorly understood processes treated in one
dimension: mass accretion efficiency, rotationally enhanced wind mass
loss, thermohaline mixing, and rotational mixing. These also impact
the composition of the envelope, and thus its radius and
$T_\mathrm{eff}$. Therefore, although not perfect, we consider the
match with the mass fractions reported by \citetalias{villamariz:05}
surprisingly satisfactory.

\section{Robustness of the models and discussion}
\label{sec:discussion}

Models of the interior evolution of stars require the use of several
poorly constrained parameters, most arising from the one-dimensional
representation of multi-dimensional phenomena (e.g., mixing,
rotation, mass loss). This remains true when modeling two stars in a
binary, with the added caveat that an even larger number of parameters
enters in the treatment of binary interactions (and in particular mass
transfer). This emphasizes the need for observational constraints and
motivated us to compare our models to the observationally well
characterized \zoph.

Accretor stars are expected in most populations of (massive) stars,
both in clusters \citep[e.g.,][]{chen:09, wang:20} and in the field
\citep[e.g.,][]{demink:11, demink:13, dorigo-jones:20}. However, these might not
obviously stand out as binary products in kinematics surveys
\citep[e.g.,][]{renzo:19walk}. Therefore, to inform the search for
accretor stars in observed samples, it is also necessary to
characterize the robustness of model predictions against numerical,
physical, and algorithmic choices.  In
\Secref{sec:single_star_uncertainties} we report on exploration of
parameter variations for each individual star in the binary and our
single star models. We
discuss parameters governing the binary evolution
\Secref{sec:bin_param}, and the consequences of the assumed SN
explosion of the companion in \Secref{sec:SN_comp}.

\subsection{Uncertainties in the single-star physics}
\label{sec:single_star_uncertainties}

\paragraph{Rotation}
Rotation is a critical ingredient of our models: it governs the
equatorial radius and thus $\omega/\omega_\mathrm{crit}$. Therefore,
by assumption, it also controls the mass transfer efficiency through
mechanical enhancement of the accretor's wind (see
\Secref{sec:bin_param}). Through Eddington-Sweet meridional
circulations, shear, and GSF instability, rotation contributes with outward mixing from the core, and more
importantly inward mixing from the surface of the accretor.  We
emphasize that the shellular approximation used in one-dimensional
stellar evolution codes might not be appropriate for
$\omega/\omega_\mathrm{crit}\simeq 1$ \citep[e.g.][]{maeder:00}, which
our accretor star reaches during RLOF.  Decreasing the diffusion
coefficient for Eddington-Sweet meridional circulations by a factor of
10 has a very small effect on the evolution of the accretor on the HR
diagram. However, the noisiness during the late RLOF phase (beyond
point B in \Figref{fig:HRD_both}) increases in amplitude, confirming
that the details of this part of the evolution are sensitive to the
treatment of rotational mixing (and its interplay with other
processes).

\paragraph{Angular momentum transport}
Our models assume a Spruit-Tayler dynamo \citep{spruit:02} for the
transport of angular momentum. Adopting the
stronger transport from \cite{fuller:19} might result
in a more efficient spin-down of the surface during RLOF, possibly
allowing for more accretion of mass.

The accretor is spun up late and from the surface inwards. It also
reaches critical rotation in contrary to the single star models, which
are initialized at ZAMS with rigid rotation. Late during the mass
transfer, even the weak core-envelope coupling of \cite{spruit:02} is
sufficient for the accretor to achieve rigid rotation (see
\Figref{fig:struct_rot}). Subsequently, the envelope spins down
because of expansion and mass loss, and more importantly the core spins
 up because of its evolutionary contraction. We expect
that following \cite{fuller:19} the core would lose more angular
momentum to the envelope, limiting its ability to spin up as it
contracts and decreasing its rotational velocity.
Nevertheless, we
expect that the accretor star will be spun up from the surface and inwards, suggesting that the differences
 with single star rotational profiles may
remain, albeit possibly smaller, because of the shorter evolutionary
time left.

\paragraph{Thermohaline mixing}
During RLOF, thermohaline mixing in the envelope becomes the dominant
mixing process. In \MESA, each mixing process is represented by its
own diffusion coefficient, and they are then summed together
\citep{paxton:11}, under the implicit assumptions that mixing
processes are independent from each other. This is typically
reasonable since locally one process dominates the mixing.  Initially,
Eddington-Sweet circulations are dominant in the accretor's envelope,
however, later on, thermohaline mixing reaches and exceeds their
diffusivity because of accretion of
chemically enriched material from the donor. If fast rotation can
physically modify thermohaline mixing processes, this could impact our
accretor models. We also computed models with enhanced efficiency of
thermohaline mixing \citep[100 times higher,
][]{schootemeijer:19}, but these proved numerically unstable when
accreting CNO-processed material.

\paragraph{Convective overshooting}
On the basis of nucleosynthesis arguments \citep[e.g.,][]{herwig:00}
and asteroseismology \citep[e.g.,][]{moravveji:16}, an exponentially
decreasing overshooting is generally considered
preferable. Nevertheless, we have also explored models with a
step-function overshooting from \cite{brott:11}. Our
exponential overshooting \citep{claret:17} was chosen to reproduce the width of the
main sequence of the models of \cite{brott:11} and, not-surprisingly,
the qualitative evolution of our fiducial model and models with step
overshooting is similar. However, adopting a step overshooting
provides a higher diffusivity at the outer edge of the core (cf.\
exponential decrease), which ultimately impacts the details of the
chemical profile at the outer edge of the core, and the morphology of
the evolutionary tracks during late RLOF.

\paragraph{Stellar winds}
\zoph\ is one of the low-luminosity O-type stars for which
\cite{marcolino:09} found a lower-than-predicted wind mass loss rate.
To address the ``weak wind problem'', we also attempted running models
with artificially decreased wind mass loss rate
\citep[e.g.,][]{renzo:17}, but these resulted in super-critically
rotating ($\omega/\omega_\mathrm{crit}>1$) post-RLOF accretor stars
with untrustworthy numerical results. The solution to the ``weak wind
problem'' is not currently known, but \cite{lucy:12} and
\cite{lagae:21} suggest that observed mass loss rates might be
underestimated, meaning that the theoretically motivated hot star
wind mass loss rate might still be appropriate to model low luminosity
O-type stars.

Our models use the \cite{vink:00, vink:01} mass-loss rate on the main
sequence, which includes the enhancement due to the bistability jump
at $T_\mathrm{eff}\simeq25\,000\,\mathrm{K}$. This results in the
dramatic increase in the surface spin down at late times (cf.\
\Figref{fig:surf_rot}). However, the mass-loss (and consequently
spin-down) enhancement at the bistability jump has recently been
questioned by \cite{bjorklund:21}. If such enhancement does not occur,
it is possible our models would retain a higher surface rotation rate,
and higher $\omega/\omega_\mathrm{crit}$. This would influence in a
similar way our single star models and our accretor, suggesting the
relative comparison between these models would still remain valid.

\paragraph{Metallicity}
Throughout this study, we assumed an initial metallicity $Z=0.01$
informed by the asteroseismology of low mass stars in
Upper-Centaurus-Lupus \citep[e.g.,][]{murphy:21}, identified as the
parent association for \zoph\ by \cite{neuhauser:20}. Moreover, we
have assumed that mass fractions of each element scale with the Solar
values \citep{grevesse:98}, which might not be appropriate especially
for massive stars \citep[e.g.,][]{grasha:21}. With these assumptions,
the initial mass fraction of $^{12}\mathrm{C}$ and $^{14}\mathrm{N}$
are lower than the values reported by \citetalias{villamariz:05} for
\zoph\ (see \Secref{sec:surf_comp}). Even though both values increase during mass transfer, our
model still slightly under-predicts them. Improved agreement could be
obtained changing the ratio of abundances to non-solar values, or by
changing the efficiency of downward rotational and thermohaline mixing
which dilutes the accreted material into the accretor's envelope.

We also ran a model identical to the one described in
\Secref{sec:best_model}, except with $Z=Z_\odot=0.0142$
\citep{asplund:09} with the same composition scaling from
\cite{grevesse:98}.  Qualitatively, the binary evolution remains
similar, with the higher metallicity stars having slightly larger
radii and cooler $T_\mathrm{eff}$ at a given luminosity. This still
produces a stable case B RLOF, however, matching the high present-day
$T_\mathrm{eff}=32\,000\pm2\,000$ of \zoph\ (e.g.,
\citetalias{villamariz:05}) requires more massive and hotter accretors
at higher $Z$ (see also \Secref{sec:bin_param}).

\subsection{Uncertainties in the treatment of mass transfer}
\label{sec:bin_param}

\paragraph{Mass transfer efficiency, $\beta_\mathrm{RLOF}$}
We regulate the accretion efficiency through the rotational
enhancement of mass loss \citep[e.g.,][]{langer:98, petrovic:05, wang:20}.  However, whether
critical rotation can effectively stop the accretion of matter is
unclear. \cite{popham:91} and \cite{paczynski:91} argued that
accretion of mass (but not angular momentum) might be possible even at
or beyond critical rotation.

During RLOF, the total amount of mass lost by the donor is
$\Delta M_\mathrm{donor} \simeq 10.6\,M_\odot$, of which only
$\Delta M_\mathrm{accretor}\simeq 3.4\,M_\odot$ are successfully
accreted by the companion. This corresponds to an overall mass
transfer efficiency
$\beta_\mathrm{RLOF}\equiv |\Delta M_\mathrm{accretor}|/|\Delta M_\mathrm{donor}| \simeq 0.32$,
although the accretion efficiency is \emph{not} constant throughout
the mass transfer \citep[e.g.,][]{vanrensbergen:06}. In our models,
the mass transfer efficiency depends on the radial and rotational
evolution of the accreting star. During RLOF, the accretor is out of
gravothermal equilibrium with significant impact on its radius and
ultimately on the amount of mass transferred and its angular
momentum. In reality, the gas stream between the two stars, the
hot-spot due to the RLOF stream hitting the accretor's surface (see
below), and the geometric distorsion of the outer layers because of
the centrifugal forces would not follow the spherical symmetry imposed
by 1D codes such as \texttt{MESA}.

While the mass-transfer efficiency, $\beta_\mathrm{RLOF}$, and
importantly its evolution need further attention, it is also
likely that this parameter depends on the details of
the system (masses, mass ratio, period, etc.). For instance, to
explain the lower mass sdO+Be binaries found by \cite{wang:21_sdOBe}
it is likely that a larger mass transfer efficiency would be
required. Conversely, \cite{petrovic:05} argued for
$\beta_\mathrm{RLOF}\simeq 0.1$ to reproduce WR+O star binaries.

Most studies, especially using rapid population synthesis tools,
typically assume a constant $\beta_\mathrm{RLOF}$ and neglect the
out-of-equilibrium phase of the accretor and how this can impact the
binary and orbital evolution. Alternatively, rapid population
synthesis can limit the accretion rate based on the thermal timescale
of the accretor (calculated from models in gravothermal
equilibrium). Based on this approach, \cite{schneider:15} found a
higher $\beta_\mathrm{RLOF}\simeq 0.7$ for a binary comparable to ours
(initially $M_1=20\,M_\odot$, $M_2=0.7M_1$ with separation
$a\simeq300\,R_\odot$), although their $\beta_\mathrm{RLOF}$ is very
sensitive to the initial mass ratio and period in this regime.

\paragraph{Specific angular momentum of accreted material}
This is an uncertain quantity and likely depends on the geometry of
the accretion process, and in particular, whether the accretion stream
through the first Lagrange point (L1) hits the accretor star directly,
or if instead an accretion disk is formed \citep[e.g.,][]{demink:13}.

We calculate the minimum distance $R_\mathrm{min}$ between the stream
coming from L1 and the accretor using the fit from \cite{ulrich:76} to
the numerical results of \cite{lubow:75}. We find
$R_\mathrm{min}\simeq 1.5\,R_\odot < R_\mathrm{accretor}$: this
suggests that the stream should hit the accretor directly, without
forming an accretion disk. Nevertheless, for the sake of numerical
stability, we assume the incoming material and the accretor's surface to
have the same specific angular momentum. This provides a slow angular
momentum accretion and consequent spin-up of the surface.

For a more physically motivated approach, we also attempted
calculations using for specific angular momentum of the accreted
material $j=\sqrt{1.7GM_\mathrm{accretor}R_\mathrm{min}}$,
representative for direct impact of the incoming stream with the
accretor \citep{lubow:75}. This is typically much larger than the
specific angular momentum of the accretor's surface. However, these
models proved numerically more unstable and providing less trustworthy
results after the accretor is spun up significantly. In general,
allowing for a faster accretion of angular momentum results in a
faster spin-up, and a lower overall mass transfer efficiency
$\beta_\mathrm{RLOF}$.

\paragraph{Specific entropy of the accreted material}
In our models, the composition of the transferred material is
determined by the structure of the donor and the mass transfer rate
calculated following \cite{kolb:90}, but we need to specify its
specific entropy when it reaches the accretor surface. We follow the
common practice of assuming the specific entropy of the incoming
material to be same as the accreting surface. The scenario justifying
this hypothesis is that during RLOF the matter is sufficiently
optically thin so that radiative processes can rapidly equalize the
entropy between the RLOF stream and the accreting surface. However,
the very large mass-transfer rates we find (cf.~\Figref{fig:MT}) might
result in optically thick flows for which this approximation might not
be appropriate.

\paragraph{Rejuvenation and core growth}
Because of the increase in mass, our accretor star is rejuvenated: its
total main-sequence lifetime is longer than the lifetime of a single
star born with the final post-RLOF mass of the accretor\footnote{But
  not significantly longer than the lifetime of a single star of its
  initial, pre-RLOF mass.}  \citep[e.g.,][]{schneider:16}. The
rejuvenation is due to the increase - in mass - of the core region,
which brings fresh nuclear fuel inwards. Our results are in agreement
with \cite{hellings:83}, while \cite{braun:95} did not find any
rejuvenation in their accretor models. We attribute this difference to
the lack of convective boundary mixing (e.g., overshooting, efficient
semiconvection, shear) in their models, which impedes the growth of
the core. In our models, the growth of the core is initially driven by
convection and overshooting, and to a lesser extent by dynamical
shear, while the off-center convective layer of \Figref{fig:D_mix} and
\Figref{fig:rho} does not contribute significantly to the inward
mixing of H-rich material and the rejuvenation itself. We cannot
exclude that in the presence of a strong shell undershooting that
convective layer would also mix efficiently with the core, enhancing
further the rejuvenation effect.

\paragraph{Initial binary parameters}
The initial donor mass $M_1$, mass ratio $q\equiv M_2/M_1$, and
period of the progenitor binary of \zoph\ cannot be directly
constrained from observations. We have explored variation in these,
and the qualitative behavior of the models is similar.  Shorter
initial periods results in larger post-RLOF orbital velocities, and
thus larger runaway velocities if the binary is disrupted at the first
SN (see \Secref{sec:SN_comp}). For example, taking $P$=75\,days (cf.\
100\,days in our fiducial model), the binary still experiences stable
case B mass transfer, but the post-RLOF orbital velocity of the
accretor is about $60\,\kms$, that is $\sim{}10\,\kms$ higher than in
our fiducial model, because of the larger orbital velocity in the
tighter binary system.

Increasing the donor mass also has a similar effect on the post-RLOF
orbital velocity of the accretor. Using $M_1=30\,M_\odot$ (cf.\
$25\,M_\odot$ in \Secref{sec:best_model}), $M_2=17\,M_\odot$, and
$P$=100\,days, we obtain a post-RLOF velocity of $65\,\kms$. However,
this produces a stripped donor of $\sim$16\,$M_\odot$ at RLOF
detachment, with stronger wind mass loss rate. Therefore this binary
is expected to widen relatively more than our fiducial model of
\Secref{sec:best_model}, slowing down the accretor. The increased mass
of the stripped donor star could also imply a lower chance of
exploding for the donor, which might instead collapse to a black-hole
(however, see \Secref{sec:SN_comp}). This would be incompatible with
the association of the pulsar PSR B1706-16 with \zoph\ \citep{neuhauser:20}.

The higher $M_1$ does not significantly change the post-RLOF total
mass of the accretor, with $M_2$ remaining about $\sim$20.5\,$M_\odot$, since
in our models accretion is regulated mostly by the spin up of the
accretor, and we do not couple the specific angular momentum of the transferred
material to the orbit or the donor's spin.

However, changing the initial mass ratio also changes the difference
between the main-sequence lifetime of the two stars, and thus how far
along the main sequence the accretor is at the onset of RLOF. The
observed position of \zoph\ on the HR diagram, particularly its
relatively high $T_\mathrm{eff}$, is difficult to reproduce assuming
initially less massive accretors (which would remain too cool even
after accreting mass), or a more equal initial mass ratio (which would
produce an accretor that is too evolved and cool at the onset of mass
transfer).

\newpage
\subsection{The explosion of the donor star}
\label{sec:SN_comp}

Throughout this study, we have assumed the ``binary SN scenario'' to
explain the runaway nature of \zoph: after the mass transfer phase,
the explosion of the donor disrupts the binary and ejects the accretor
at roughly its pre-explosion orbital velocity
\citep[e.g.,][]{blaauw:61, eldridge:11, renzo:19walk}. This fate occurs to the
majority of massive binary systems, and \zoph\ might be the best
example of it \citep[e.g.,][]{blaauw:52, blaauw:61,
  hoogerwerf:00}. \cite{neuhauser:20} suggested not only the companion
successfully exploded producing the pulsar PSR B1706-16 and ejecting
\zoph, but also that the explosion produced radioactive
$^{60}\mathrm{Fe}$ which polluted Earth.

From kinematic and orbital considerations they estimated the pulsar
received a natal kick of $253\pm54\,\kms$, which would be sufficiently
large to unbind the binary which has
$v_\mathrm{orb}=\sqrt{G(M_1+M_2)/a}\simeq 135\,\kms$ at the end of our
binary simulation (blue diamond in \Figref{fig:HRD_both}), and this
will decrease further in the remaining time to the donor's
core-collapse \citep{kalogera:96, tauris:98, tauris:15}.

The SN ejecta mass would depend on the post-RLOF wind mass loss of our
donor star, which is uncertain \citep[see also][]{renzo:17, vink:17,
  gilkis:19, sander:20}. At the end of our binary evolution simulation, our
stripped donor is $\sim{}9.4\,M_\odot$, with a surface H fraction of
$X\lesssim0.2$ for a layer of $\Delta M \simeq 2.5\,M_\odot$.  Its
wind mass-loss rate is $\sim10^{-5}\,M_\odot \mathrm{yr^{-1}}$,
calculated assuming the empirical Wolf-Rayet wind mass loss
prescription from \citet[][see also \Figref{fig:MT}]{nugis:00}. We
expect that the donor will explode in a H-free type Ib
supernova. Although our stripped donor is rather massive, recent
studies hints at a higher ``explodability'' of donor stars in binary
systems \citep[e.g.,][]{schneider:21, laplace:21, vartanyan:21}.

We have neglected the impact of the explosion on the structure of the
accretor star. At the time of the explosion, the accretor subtends a
solid angle $\sim{}R^2/a^2\simeq 2\times10^{-3}$\,steradians with $R$
the accretor radius and $a$ the binary separation. We neglect the
post-RLOF wind-driven orbital widening for this estimate.  The blast
wave will hit the accretor causing mass loss -- directly via ablation
and by injecting energy in the envelope, inflating it and enhancing
its wind \citep{wheeler:75, tauris:98, podsiadlowski:03, hirai:18, ogata:21}.
Because of the SN shock, the just ejected new runaway star might
appear bloated and redder (long before it overtakes the slowing SN
remnant). The impact of this brief out of thermal equilibrium phase on
the stellar spin should be investigated further.

Using 2D hydrodynamic simulations of the star-SN ejecta interactions
in close binaries ($a\lesssim 60\,R_\odot$,
cf. $a\gtrsim 343\,R_\odot$ in our fiducial binary model),
\cite{hirai:18} found that the companion star recovers its
pre-explosion luminosity and effective temperature within a few years
to decades. This result was later corroborated for a wider range of
parameters by \cite{ogata:21}. The amount of mass removed by the SN
shock is $\lesssim10^{-2}\,M_\odot$.  The SN ejecta might also pollute
the surface of the runaway by depositing processed nuclear material
\citep[e.g.,][]{przybilla:08, suda:21}. However, for the large final
separation of our model, little pollution is expected and enhanced
mass loss and inward mixing might quickly dilute any signature below
detectable levels.

\section{Summary \& conclusions}
\label{sec:conclusions}

The impact of mass transfer on the structure and evolution of
accretor stars in massive binaries has received relatively little
attention. To investigate this, we have performed \texttt{MESA}
calculations of massive binaries evolving two coupled stars
simultaneously.

As a first application, we focused on finding a model in which the
accretor properties are in qualitative agreement with observations of
the nearest O-type star to Earth. This is the runaway star \zoph, which has
long been suggested to be a former accretor star ejected from a binary
at the core-collapse of the donor star \citep[binary SN
scenario,][]{blaauw:61}. However, our models are also informative for
the generic population of massive stars accreting in binaries.

\subsection{Reproducing \zoph}

We found that the main features of \zoph\ can be
reasonably well reproduced using standard non-fine-tuned stellar physics
assumptions for the treatment of mass transfer, chemical mixing, and
rotation. Our choices are described in \Secref{sec:methods} and
Appendix~\ref{sec:software}.

Our fiducial model is a binary starting
with $M_1=25\,M_\odot$, $M_2=17\,M_\odot$, and $P=100$\,days at
metallicity $Z=0.01$ (see \Secref{sec:best_model}). This binary
experiences stable thermal-timescale Roche lobe overflow after the end
of the main sequence of the donor (i.e., case B type mass transfer).

The accretor compares well with observations of \zoph\ about
$1.5-2$\,Myrs after the end of mass transfer, corresponding to the
remaining donor's lifetime at the end of our simulations plus the
kinematic age of \zoph. Specifically, the position on the HR diagram
(cf.~\Figref{fig:HRD_both}), the runaway space velocity (estimated
based on the accretor's orbital velocity), the surface composition and
rotational velocity (cf.~\Tabref{tab:surf_prop}) are in the right
ballpark.

Our model of \zoph\ differs significantly from previous studies: in
contrast with the accretor models of \cite{vanrensbergen:96}, in our
model the $^{14}\mathrm{N}$- and $^4\mathrm{He}$-rich surface
composition is not the result of pure outward rotational mixing.  In
addition \referee{and more importantly}, \referee{CNO-enriched matter}
transferred from the layers above the core of the donor star
\referee{contributes to the surface composition. This material is}
mixed from the surface inwards into the accretor by meridional
circulations and, more importantly, thermohaline mixing.  Thus, the
present day surface mass fractions of \zoph\ constrain the mass
transfer efficiency and mixing in the accretor. Our results suggest
\zoph\ should not be used to calibrate models of rotational mixing in
single star models.

We emphasize that the surface composition alone would not be a
smoking-gun of a binary evolution history, especially given the large
uncertainties in the treatment of rotation and mixing in stellar
evolution models. In our models, the accretion of $^{14}\mathrm{N}$
from the donor star allows the accretor star to be simultaneously fast
rotating and $^{14}\mathrm{N}$-rich. Alternative scenarios where
\zoph\ evolved as a single, fast-rotating star require ad-hoc
explanations for the runaway velocity, and have been shown by
\citetalias{villamariz:05} to struggle in reproducing surface mass
fractions, apparent age, mass, and rotation rate simultaneously.

The surface rotation rate of the accretor post-mass-transfer is always
higher than the rotation rate of single stars initialized with
half-critical rotation, but might still be on the low side compared to
\zoph. However, the wind spin down might be overestimated in our
models (weak wind problem, cf.\ \citealt{marcolino:09, lucy:12, lagae:21}).

We also tested the
robustness of our fiducial model against variations in the initial parameters
and algorithmic representation of physical phenomena, discussed in
\Secref{sec:discussion}. Less massive accretors remain too cool
throughout the evolution to be compatible with \zoph, and initial mass
ratios closer to unity lead to a more evolved accretor at the onset of
mass transfer, again resulting in too cool temperatures. Increasing
the donor's initial mass might result in stripped stars unlikely to
form a neutron star in their final SN explosion.

\subsection{Accretors are not single rotating stars}

Our models also highlight some general differences between accretors in massive
binaries and stars evolving as single throughout their life. These
might be important for several
sub-fields of astrophysics, including asteroseismology, stellar
populations, and time-domain and gravitational waves observations.

The first notable difference we find is the internal rotation
profile. Single rotating stars are usually initialized as rigid
rotators at birth, and throughout their evolution they spin down due
to wind mass loss. Conversely, accretors are spun up later during
their main sequence and from the surface inward. Moreover,
for single stars, the maximum rotation rate, that is the one assumed
at the beginning of the evolution, is poorly understood theoretically
and observationally \citep[e.g.,][]{ramirez-agudelo:13,
  ramirez-agudelo:15}. Conversely, accretors in binaries reach
critical rotation $\omega/\omega_\mathrm{crit}\simeq 1$
\citep[e.g.,][]{packet:81}. The later spin-up and higher achieved
rotation rate allow the accretor star to remain a fast rotator until
the end of its main sequence.

The angular momentum accreted at the surface of the accretor is
transported into the core (by the Tayler-Spruit dynamo in our
simulations). This results in a much faster rotating helium core at
the end of the main sequence compared to single stars. Such a fast
spinning helium core has potential implications for the final
explosion and the resulting compact object born from the accretor star
in an interacting binary system.

Finally, in our models, the accretion of mass leads to rejuvenation
and also the formation of a off-center convective layer above the
main-sequence core (cf.\ \Figref{fig:D_mix}). The latter ultimately
results in a sharper density drop at the core edge, and a flatter
density profile close to the end of the main sequence (cf.\
\Figref{fig:rho}). If physical,
the presence of such a convective layer could in principle be probed
using asteroseismology. Depending on how the accretor (and the binary)
evolves in the future, this difference could be crucial in determining
the outcome of common
envelope events between massive stars and compact objects.\\

Improving our understanding of the evolution of the initially less
massive stars in massive binary systems is crucial for the upcoming
large surveys, stellar kinematics, and for the understanding of the
evolution of gravitational-wave progenitors in isolated
binaries. Although presently single, the nearest O-type star to Earth,
\zoph, can be used as an anchor point for the modeling of
accretors. Our models demonstrate that a broad agreement with
observations can be achieved with standard stellar evolution
assumptions.

\referee{However, our case study should be extended to understand more
  generally the evolution of accretor stars.} Future efforts should
\referee{explore the parameter space for mass transfer and mixing processes,
  and consider} a wider mass, period, mass ratio, and metallicity
range to investigate the impact of binary evolution on the life,
explosion, and after-life of the secondary stars in massive binary
systems.

\software{
  \MESA\ \citep{paxton:11,paxton:13,paxton:15,paxton:18,paxton:19},
  \texttt{mesaSDK} \citep{mesasdk},
  \texttt{ipython/jupyter} \citep{ipython},
  \texttt{matplotlib} \citep{matplotlib},
  \texttt{mesaPlot} \citep{mesaplot},
  \texttt{NumPy} \citep{numpy}.
}

\acknowledgements{We are grateful to E.~Zapartas, A.~Jermyn,
  M.~Cantiello, R.~Neuh\"auser, B.~D.~Metzger, S.~Oey, and S.~Justham
  for helpful discussions and feedback. We also thank P.~Marchant for
  helpful guidance with the newe superadiabaticity reduction
  capabilities in \texttt{MESA}. Support for this work was provided by
  NASA through the NASA Hubble Fellowship Program grant
  \#HST-HF2-51457.001-A awarded by the Space Telescope Science
  Institute, which is operated by the Association of Universities for
  Research in Astronomy, Inc., for NASA, under contract NAS5-26555.}

\appendix

\section{\texttt{MESA} setup}
\label{sec:software}

We use \code{MESA} version 15140 to compute our models.  The
\code{MESA} equation of state (EOS) is a blend of the OPAL \citet{Rogers2002}, SCVH
\citet{Saumon1995}, PTEH \citet{Pols1995}, HELM \citet{Timmes2000},
and PC \citet{Potekhin2010} EOSes.

OPAL \citep{Iglesias1993, Iglesias1996} provides the main radiative
opacities, with low-temperature data from \citet{Ferguson2005} and the
high-temperature from \citet{Buchler1976}. Electron conduction
opacities are from \citet{Cassisi2007}.

Nuclear reaction rates are a combination of rates from NACRE
\citep{Angulo1999}, JINA REACLIB \citep{Cyburt2010}, plus additional
tabulated weak reaction rates \citet{Fuller1985, Oda1994,
  Langanke2000}. Screening is included via the prescription of
\citet{Chugunov2007}.  Thermal neutrino loss rates are from
\citet{Itoh1996}. We use a
22-isotope nuclear network (\texttt{approx\_21\_plus\_cr56}).

The inlists, processing scripts, and model output are available at~\url{https://doi.org/10.5281/zenodo.4701565}.

\section{Resolution tests}
\label{sec:res_tests}

\begin{figure*}[htbp]
  \centering
  \includegraphics[width=\textwidth]{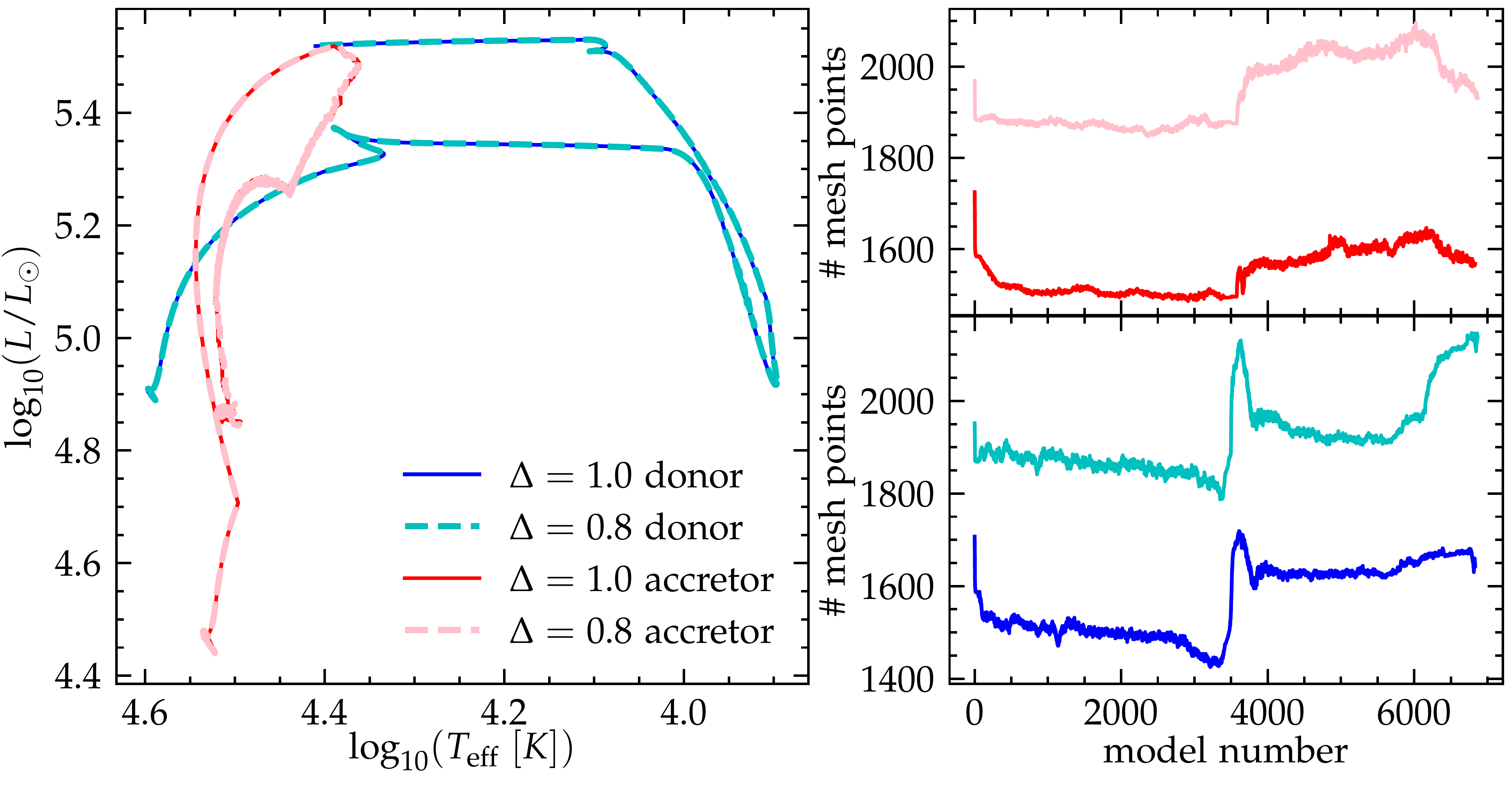}
  \caption{Left: HR diagram comparison for our fiducial binary model varying
  the number of mesh points. We only show the evolution until our definition
  of RLOF detachment. Right: number of mesh points as a
  function of timestep number. In both panels, the blue/cyan tracks show the donor stars, the
red/pink tracks show the accretor. Thicker dashed lines correspond to
the models at higher resolution (i.e., lower $\Delta$ which indicates
the value of \texttt{mesh\_delta\_coeff}).}
\label{fig:sp_test}
\end{figure*}

We extensively checked the numerical convergence of our stellar
evolution calculations with increasing number of mesh
points. \Figref{fig:sp_test} shows that all the main features described
here do not vary when increasing the spatial resolution by increasing
the number of mesh points (i.e.,
decreasing \texttt{mesh\_delta\_coeff}). The right panel shows the
number of mesh points for the accretor (top) and donor (bottom) as a
function of the model number (akin to an arbitrary time
coordinate). About $\sim$7000 \texttt{MESA} timesteps are used to compute
the binary evolution. The higher resolution run has $\sim 20\%$ more
mesh points. The left panel shows the evolution on the HR diagram until the
detachment of the binary for the two accretor models (pink/red) and
the two donor models (blue/cyan).

Similarly, we tested the numerical convergence with decreasing
timestep size. This can be done decreasing the parameter
\texttt{mesh\_time\_coeff}. However, we were unable to successfully
compute models at higher temporal resolution. Partial results show a
good agreement with our fiducial model until \texttt{MESA} becomes
unable to find a satisfying numerical solution to the stellar
structure equations (typically
during RLOF). Lower temporal resolution models showed a similar
qualitative agreement but increased noisiness during the late RLOF
phase. For our fiducial model the adaptive timestep size never exceeds
$10^{3.8}$\,years with typical pre-RLOF timesteps of the order of $10^{3.2}$\,years
and sub-decade (occasionally sub-year) during RLOF. The main factor limiting the timestep
sizes is the change of surface angular momentum in both stars during
the mass transfer.

\section{Internal composition profile evolution and mixing}
\label{sec:X_fig}

\begin{figure*}[htbp]
  \centering
  \includegraphics[width=\textwidth]{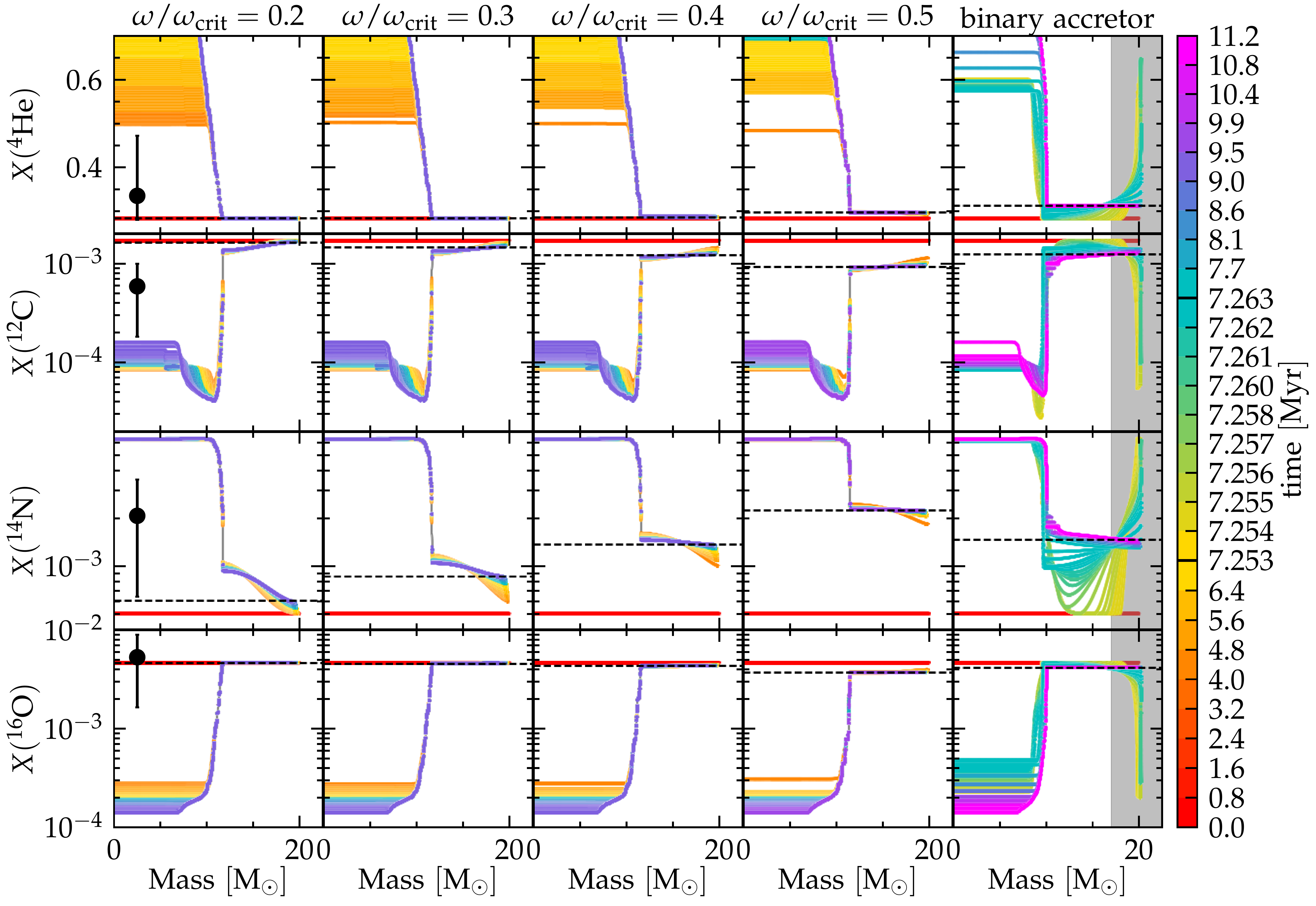}
  \caption{Same as \Figref{fig:n14}, but for
    \referee{$^{4}\mathrm{He}$ (top row), $^{12}\mathrm{C}$ (second
      row), $^{14}\mathrm{N}$ (third row, identical to \Figref{fig:n14}),
      and $^{16}\mathrm{O}$ (bottom row). We emphasize the linear
    scale on the vertical axis in the top row, while all others have a
    logarithmic scale.} The first four panels show single rotating
    stars of initially $20\,M_\odot$, the rightmost panels show the
    accretor in our fiducial binary. The red lines mark the initial
    mass fractions, black dashed lines indicate the TAMS surface mass
    fraction, and the black error bars in the first column indicate
    the surface composition of \zoph\ inferred by
    \citetalias{villamariz:05} using the surface H mass fraction from
    our model.}
  \label{fig:composition_huge}
\end{figure*}

\Figref{fig:composition_huge} compares the internal evolution of the
composition profile of single rotating stars with our accretor model.
We show mass fractions of \referee{$^{4}\mathrm{He}$}, $^{12}\mathrm{C}$, and
$^{16}\mathrm{O}$ to complement the mass fraction of $^{14}\mathrm{N}$
shown in \Figref{fig:n14}, and reproduced also in \referee{third row} of
\Figref{fig:composition_huge}. \referee{The black error bars in the
  leftmost panel correspond to the abundances reported by
  \citetalias{villamariz:05} assuming a surface hydrogen mass fraction
  $X_\mathrm{surf}=0.67801$ (cf.\ \Tabref{tab:surf_prop}).}

\begin{figure*}[htbp]
  \centering
  \begin{interactive}{animation}{composition_accretor.mp4}
    \includegraphics[width=0.5\textwidth]{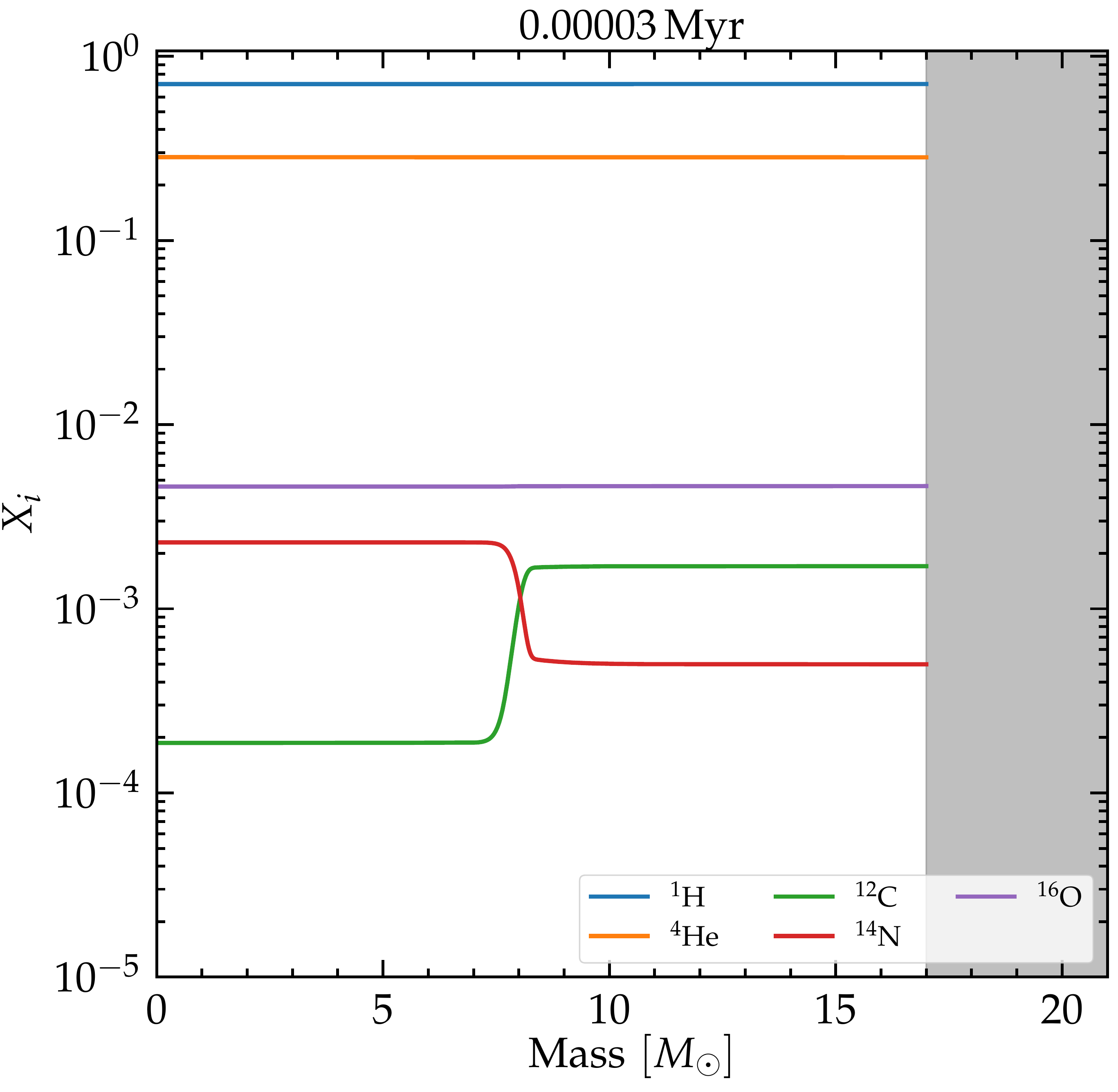}
  \end{interactive}
  \caption{Composition profile of our fiducial accretor model at the
    beginning of the simulation (ZAMS). The animation shows the entire
    main-sequence evolution (see physical time at the top). The number of frames per unit time is not constant, and greatly
    increases during the RLOF phase, when smaller timesteps are needed.}
  \label{fig:composition_movie}
\end{figure*}

\referee{ \Figref{fig:composition_movie} shows the composition profile
  of $^1\mathrm{H}$, $^{4}\mathrm{He}$, $^{12}\mathrm{C}$,
  $^{14}\mathrm{N}$, and $^{16}\mathrm{O}$ for our fiducial accretor
  model shortly after ZAMS. The associated animation shows how these
  evolve because of nuclear processes in the core, accretion at the
  surface, and mixing both inward (during RLOF because of thermohaline
  and rotation) and outward (rotation).}

\begin{figure*}[htbp]
    \centering
  \begin{interactive}{animation}{mix_movie.mp4}
    \includegraphics[width=0.5\textwidth]{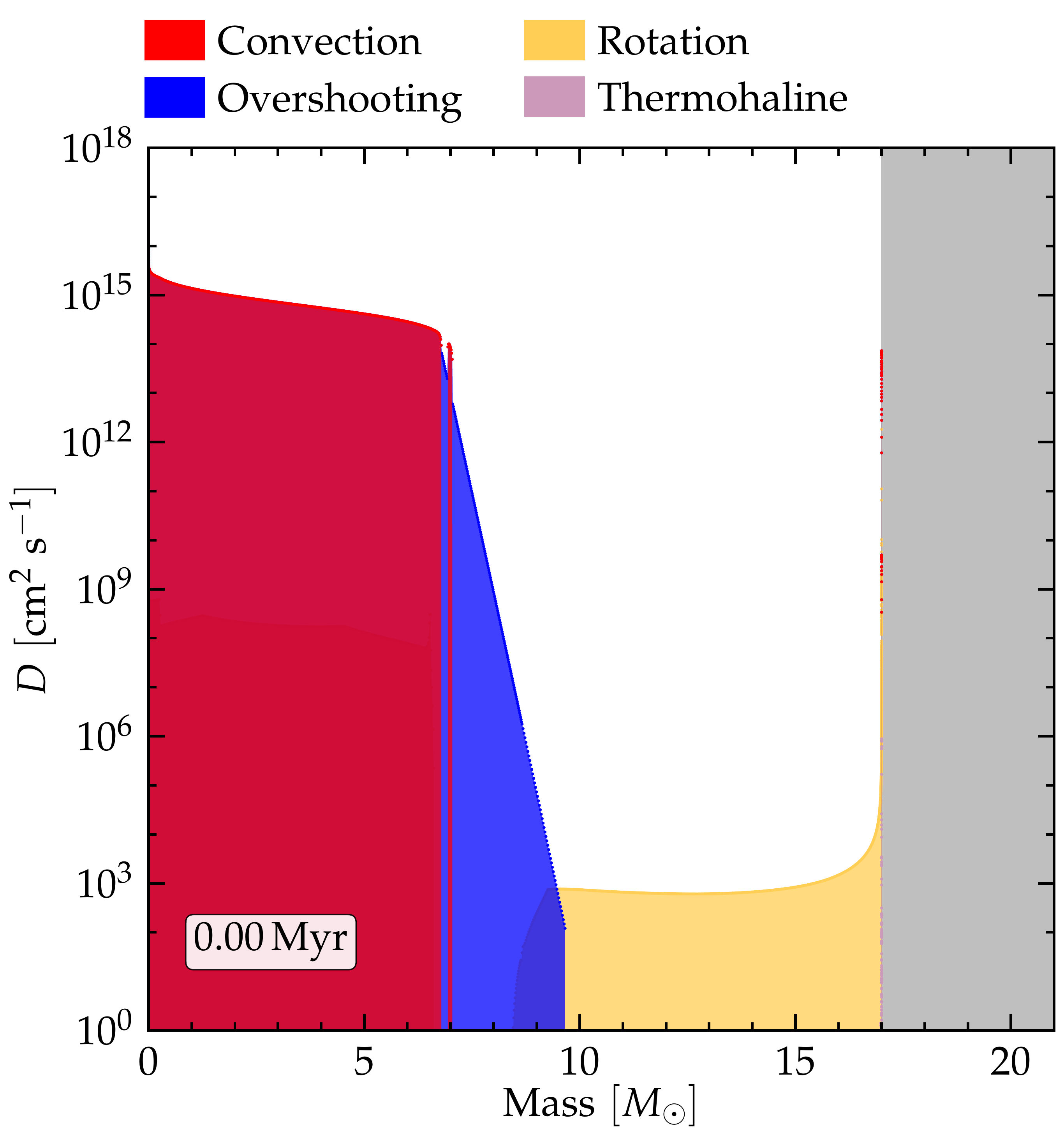}
  \end{interactive}
  \caption{Diffusion coefficients for convection (red), overshooting
    (blue), thermohaline (pink), and rotational mixing (yellow). The
    latter includes meridional circulations, secular and dynamical
    shear, and GSF. Semiconvection is not shown for clarity. The
    associated animation shows how these evolve with time until
    TAMS. The number of frames per unit time is not constant, and
    greatly increases during the RLOF phase, when smaller timesteps
    are needed.}
  \label{fig:mix_movie}
\end{figure*}

\referee{\Figref{fig:mix_movie} shows the diffusion coefficients for
  various mixing processes (cf.\ \Figref{fig:D_mix}) at ZAMS. The
  associated animation shows the evolution of these mixing
  coefficients throughout our simulation until TAMS.}

\bibliographystyle{aasjournal}
\bibliography{./zeta_ophiuchi.bib}

\end{document}